\documentclass[11pt]{article}
\usepackage{geometry}  % Flexible and complete interface to document dimensions
\usepackage[T1]{fontenc}  % Standard package for selecting font encodings
\usepackage[utf8x]{inputenc}  % Accept different input encodings
\usepackage{lineno}  % Line numbers on paragraphs
\usepackage{fancyhdr}  % Extensive control of page headers and footers in LaTeX2ε
\usepackage{enumitem}  % Control layout of itemize, enumerate, description
\usepackage{hyperref}  % Extensive support for hypertext in LaTeX
\usepackage{titling}  % Control over the typesetting of the \maketitle command
\usepackage[numbers]{natbib}  % Flexible bibliography support
\usepackage{mathtools}  % Mathematical tools to use with amsmath
\usepackage{titlesec}  % Select alternative section titles
\usepackage{lastpage}  % Reference last page for Page N of M type footers
\usepackage[english]{babel}  % Multilingual support for Plain TeX or LaTeX
\usepackage{amsmath}  % AMS mathematical facilities for LaTeX
\usepackage{amsfonts}  % TeX fonts from the American Mathematical Society
\usepackage{amssymb}  % Additional symbols from American Mathematical Society
\usepackage{wasysym}  % LaTeX support file to use the WASY2 fonts
\usepackage{bbm}  % "Blackboard-style" cm fonts
\usepackage{array}  % Extending the array and tabular environments
\usepackage{xr}  % References to other LaTeX documents
\usepackage{verbatim} % Reimplementation of and extensions to LaTeX verbatim
\usepackage{float} % Improved interface for floating objects
\usepackage{color}% For colored text
\usepackage{algorithm, caption}
\usepackage[noend]{algpseudocode}
\usepackage{xpatch} % helps to space algorithmic environment 
\usepackage{xspace} % helps to space algorithmic environment 
\usepackage{booktabs}
\usepackage[table,xcdraw]{xcolor}
\usepackage{multirow}

\usepackage{subcaption}
\usepackage{graphicx}
\usepackage[normalem]{ulem}
\usepackage{soul}

\hypersetup{%
  pdfborderstyle={/S/U/W 1}% border style will be underline of width 1pt
}

\setcitestyle{round,numbers}
\makeatletter
\def\BState{\State\hskip-\ALG@thistlm}
\makeatother

\titleformat{name=\section}{\normalfont\LARGE\bfseries}{}{0pt}{}
\titleformat{name=\subsection}{\normalfont\Large\bfseries}{}{0pt}{}
\titleformat{name=\subsubsection}{\normalfont\large\bfseries}{}{0pt}{}

\newcommand{\email}[1]{\href{mailto:{#1}}{{#1}}}

\newcommand{\keywords}[1]{\textbf{Keywords}: {#1}}

\newcommand{\optincludegraphics}[2][]{}
\newcommand{\optinput}[1]{}
\newtagform{brackets}{[}{]}
\usetagform{brackets}

\newcommand{\thejournal}[1]{Magnetic Resonance in Medicine}

  \newcommand{\textk}[1]{\textcolor{black}{#1}}

\title{Fully Self-Gated Whole-Heart 4D Flow Imaging from a Five-Minute Scan}
\immediate\write18{texcount -sum=1,1,1,1,1,1,1 -merge -incbib -dir -utf8 \jobname.tex > \jobname.wcTotal }
\immediate\write18{texcount -sub=none -merge -incbib -dir -utf8 \jobname.tex | grep _main_ | grep -oE '[0-9]+' | python -c "import sys; print(sum(int(l) * w for l, w in zip(sys.stdin, [1, 1, 0, 0, 0, 0, 0])))" > \jobname.wcManuscript }
\immediate\write18{texcount -sub=none -merge -incbib -dir -utf8 \jobname.tex | grep Abstract | grep -oE '[0-9]+' | python -c "import sys; print(sum(int(l) * w for l, w in zip(sys.stdin, [1, 1, 0, 0, 0, 0, 0])))" > \jobname.wcAbstract }

\newcommand{\wcTotal}{\clearpage{\noindent\large{\bf Detailed Word Count} (not to be included for submission)}\verbatiminput{\jobname.wcTotal}}
\newcommand{\wcManuscript}{\input{\jobname.wcManuscript}}
\newcommand{\wcAbstract}{\input{\jobname.wcAbstract}}

\begin{document}

% ======================================================================
\begin{titlepage}
	\begin{center}
		\sf{To Appear in Magnetic Resonance in Medicine}\\
		\bigskip
		{\noindent\LARGE\bf \thetitle}
	\end{center}
	\bigskip

% : Insert author names, affiliations and corresponding author email
% : (do not include titles, positions, or degrees).
%FIXME
\begin{center}\large
	Aaron Pruitt,\textsuperscript{1}
	Adam Rich,\textsuperscript{1,2}
	Yingmin Liu,\textsuperscript{3}
	Ning Jin,\textsuperscript{4}
	Lee Potter,\textsuperscript{2,3}
    \textk{Matthew Tong,\textsuperscript{5}}
	\textk{Saurabh Rajpal,\textsuperscript{5}}
	Orlando Simonetti,\textsuperscript{3,5,6}
	Rizwan Ahmad\textsuperscript{*1,2,3}
\end{center}

\bigskip

\noindent
%FIXME
\begin{enumerate}[label=\textbf{\arabic*}]
\item Biomedical Engineering, The Ohio State University, Columbus OH, USA
\item Electrical and Computer Engineering, The Ohio State University, Columbus OH, USA
\item Davis Heart \& Lung Research Institute, The Ohio State University, Columbus OH, USA
\item Cardiovascular MR R\&D, Siemens Medical Solutions USA Inc., Columbus OH USA
\item Internal Medicine, The Ohio State University, Columbus OH, USA
\item Radiology, The Ohio State University, Columbus OH, USA
\end{enumerate}

\bigskip

% : Use the dagger symbol to denote a single equal contribution authorship.
% : Multiple equal-contribution authorship may be included in the acknowledgments.
%FIXME
%\textbf{{†}}: These authors contributed equally to this work.

% : Use the asterisk to denote corresponding authorship.
% : Provide email address in note below.
%FIXME
\textbf{*} Corresponding author:

\indent\indent
\begin{tabular}{>{\bfseries}rl}
Name		& Rizwan Ahmad													\\
Department	& Biomedical Engineering													\\
Institute	& The Ohio State University														\\
Address 	& 460 W 12th Ave, Room 318														\\
			& Columbus OH 43210, USA														\\
E-mail		& \email{ahmad.46@osu.edu}											\\
\end{tabular}

\vfill

% ======================================================================
% : set word count results (+++ must be included, --- must be excluded)
% 	+++ introduction, theory, methods, results, discussion, conclusion,
%		appendix, 
% 	--- title page, abstract, figure captions, tables, table captions,
%		references, revision markings
% : first argument is the manuscript word count
% : second argument is the abstract word count
% : to use `texcount` results, use '%TC:ignore'/'%TC:endignore' directives.
% : \wcManuscript and \wcAbstract should perform the correct word count.
%\textk{\wordcount{\wcManuscript}{\wcAbstract}

\indent\indent
\begin{tabular}{>{\bfseries}rl}
	Manuscript Word Count:	& \textk{5217}\\
	Abstract  Word Count:	& \textk{183} \\
\end{tabular}

% : display detailed word count
%FIXME
%\textk{\wordcount{\wcManuscript}{\wcAbstract}

\end{titlepage}
%TC:endignore
% ======================================================================

% ======================================================================
% ======================================================================
\pagebreak
% ======================================================================
% ======================================================================

% ======================================================================
%TC:break Abstract
\begin{abstract}
%FIXME

\noindent
\textbf{Purpose:} To develop and validate an acquisition and processing technique that enables fully self-gated 4D flow imaging with whole-heart coverage in a fixed five-minute scan. \\
\textbf{Theory and Methods:} The data are acquired continuously using Cartesian sampling and sorted into respiratory and cardiac bins using the self-gating signal. The reconstruction is performed using a recently proposed Bayesian method called ReVEAL4D. \textk{ReVEAL4D is validated using data from eight healthy volunteers and two patients and compared with compressed sensing technique, L1-SENSE.} \\
\textbf{Results:} \textk{Healthy subjects $-$} Compared to 2D phase-contrast MRI (2D-PC), flow quantification from ReVEAL4D shows no significant bias. In contrast, the peak velocity and peak flow rate for L1-SENSE are significantly underestimated. Compared to traditional parallel MRI-based 4D flow imaging, ReVEAL4D demonstrates small but significant biases in net flow and peak flow rate, with no significant bias in peak velocity. All three indices are significantly and more markedly underestimated by L1-SENSE. \textk{Patients $ - $ Flow quantification from ReVEAL4D agrees well with the 2D-PC reference. In contrast, L1-SENSE markedly underestimated peak velocity.}   \\
\textbf{Conclusions:} The combination of highly accelerated five-minute Cartesian acquisition, self-gating, and ReVEAL4D enables whole-heart 4D flow imaging with accurate flow quantification.
\end{abstract}
% ======================================================================
% : set search-engine keywords (3 to 6)
\bigskip
\keywords{4D flow, phase-contrast, self-gating, Cartesian, Bayesian, CMR}

%TC:break _main_
\pagebreak

\section{INTRODUCTION}

Cardiovascular Magnetic Resonance (CMR) is a powerful imaging tool that provides comprehensive evaluation of the cardiovascular system, including hemodynamic assessment, using phase-contrast MRI (PC-MRI) \cite{Nayak2015}. For routine clinical examinations, PC-MRI is generally limited to planar imaging and encodes only a single component of the blood velocity vector. While effective in answering certain clinical questions, PC-MRI suffers from limited coverage, requires precise slice prescription, and may not adequately capture complicated or pathological flow dynamics~\cite{DaSilveira2017, Nordmeyer2013}. Consequently, 4D flow imaging has emerged as a more comprehensive alternative for measuring hemodynamics \cite{Markl2012}. Extending the principles of PC-MRI, 4D flow imaging provides full volumetric and temporally resolved mapping of the three-dimensional velocity vector, offering an advantage over PC-MRI with regard to anatomical coverage and hemodynamic visualization. Post-processing enables retrospective interrogation of arbitrary slice planes, long after the patient is removed from the magnet. Advanced hemodynamic parameters can also be calculated which may carry additional prognostic value \cite{Markl2016,Markl2010,Owen2016}. As a result, 4D flow imaging has repeatedly been demonstrated to provide clinical information in the diagnosis and assessment of patients with congenital
\cite{SSVasanawala2015,Valverde2010,Geiger2011} and non-congenital \cite{DaSilveira2017,Hope2007,JJWestenberg2008} heart disease.

Despite its advantages, clinical adoption of 4D flow imaging has been hampered by prohibitively long acquisitions times. Recent developments in data acquisition and reconstruction, however, have afforded substantial improvements in achievable acceleration, bringing 4D flow imaging a step closer to routine practice. Expanding upon conventional parallel imaging methods \cite{Pruessmann1999,Griswold2002}, Bollache et al. demonstrated a five-fold acceleration for 4D flow imaging of the aorta using PEAK-GRAPPA, achieving a nominal two-minute acquisition time without respiratory gating \cite{Bollache2018}. Leveraging compressed sensing (CS), Ma et al. coupled randomized sampling with a wavelet penalty to recover 4D flow images of the aorta with 7.7-fold acceleration \cite{Ma2019}. Similarly, Knoblock et al. exploited complex difference sparsity in conjunction with temporal principal component analysis yielding an eight-fold acceleration for imaging the carotid bifurcation \cite{Knobloch2013}. Rich et al. used Bayesian modeling to enable 4D flow imaging of the aortic valve in a single breath-hold, achieving acceleration rates as high as 27   \cite{Rich2019}. 

Most contemporary 4D flow imaging techniques are based on free-breathing segmented acquisition, requiring real-time monitoring of physiological motion states. Electrocardiogram (ECG) is typically recorded and used to retrospectively synchronize and populate k-space data across multiple cardiac phases. The respiratory motion is compensated using navigator-based respiratory gating~\cite{Ehman1989}. Within the past several years, significant advancements have been made towards accelerated, free-running (e.g., non-gated and free-breathing) dynamic volumetric imaging \cite{Feng2016,Han2017,Feng2018}. Generally, these methods share several common characteristics, including golden angle incremented sampling \cite{Winkelmann2006}, CS-inspired reconstruction \cite{Lustig2007}, and self-gating ~\cite{Larson2004,Stehning2005}. Unlike segmented acquisition with prospective gating, data-driven self-gating provides the flexibility to distribute the data across arbitrary numbers of cardiac and respiratory phases, which enables the temporal resolution, net acceleration, gating efficiency, and extent of respiratory artifacts to be adjusted post-acquisition. 

Early work combining self-gating with 4D flow imaging was proposed by Uribe et al. \cite{Uribe2009}, whereby interleaved self-gating readouts were analyzed in real-time for respiratory gating in a way analogous to navigator-based gating. More recently, Cheng et al. proposed XD-FLOW \cite{Cheng2017} which coupled Butterfly navigators \cite{Lustig2007b} with ECG for retrospective multi-dimensional reconstruction of contrast-enhanced 4D flow imaging in pediatric patients with a mean acquisition time of approximately nine minutes. Bastkowski et al. \cite{Bastkowski2018} were the first to achieve fully self-gated 4D flow imaging in the aorta using a golden angle incremented stack of spirals sampling within a fixed 15-minute exam. Finally, Walheim et al. \cite{Walheim2019} proposed locally low rank reconstruction of 5D flow datasets using pseudo-spiral Cartesian sampling and respiratory self-gating to achieve aortic imaging from a four-minute acquisition. Despite these recent developments, contrast-free, rapid 4D flow imaging with whole-heart coverage and isotropic resolution remains a challenge. 

In this work, we propose a comprehensive whole-heart 4D flow imaging framework. The integration of the following three features comprise the technical innovation: (i) a Cartesian sampling pattern which promotes incoherence and uniformity by pseudo-random advancement in both angular and radial directions, (ii) cardiac and respiratory self-gating signals derived from a Cartesian sampling pattern for 4D flow, and (iii) reconstruction via the ReVEAL4D algorithm~\cite{Rich2019} that uses Bayesian modeling to jointly reconstruct velocity encodings. The result is a self-gated, contrast-free whole-heart 4D flow imaging that can be completed in a fixed five-minute acquisition time and provides accurate flow quantification with respect to conventional phase-contrast and navigator-gated 4D flow imaging.

\section{THEORY}
	\subsection{Incoherent Cartesian Sampling}
    In this section, we describe a pseudo-random Cartesian sampling pattern for 4D flow imaging. %The sampling pattern can be readily implemented on the scanner without using lookup tables or solving an optimization problem. 
    Given a $k_y$-$k_z$ k-space grid with matrix size $n_y \times n_z$, readout indices are initially defined in a polar coordinate space parameterized by azimuthal angle, $\theta$, and radius, $r$. Beginning with arbitrary initialization, angular indices are incremented by the golden ratio, ($g_\theta=(\sqrt{5} +1)/2$), yielding
	\begin{equation}
	\label{eq:ang}
	\theta(e, i+1) = 2\pi\left(2-g_\theta\right)\left(i+\frac{e-1}{E}\right),
	\end{equation}
	where $i$ denotes the current sample index, $E$ is the number of velocity encodings, and $e=1,2,\cdots, E$ is the velocity encoding index. Radial indices are advanced according to a second, mutually irrational number, $g_r$, yielding
	\begin{equation}
	\label{eq:rad}
	r(i+1) = {	\text{mod} \Big{(} r(i)+{g_{r}}r_{\max},\text{ } r_{\max}\Big{)}	},
	\end{equation}
	where $r_{\max}= \max\Big(\Big\lfloor\frac{n_y}{2}\Big\rfloor, \Big\lfloor\frac{n_z}{2}\Big\rfloor\Big)$ represents the maximum allowed radius and $g_r$ is set to $\sqrt[3]{35}$. Consequently, both angular and radial indices evolve in a cyclic and mutually incoherent manner. Variable density is enabled through transformation of the radial indices obtained in Eq.~\ref{eq:rad} according to
	\begin{eqnarray}
	\label{eq:scale}
	\tilde{r}(i) = \mathcal{S}\{r(i)\}, \text{   where}
	\\
	\mathcal{S}\{r\} := (r+c)^d - c^d.
	\end{eqnarray}	
	Here, $\mathcal{S}$ represents the tuneable, nonlinear scaling operation that controls sampling density in k-space and is parameterized by the user-defined constants $c$ and $d$, \textk{here empirically set to $\sqrt[4]{(n_y\times n_z)/4}$ and $1.5$, respectively.} 
	
\begin{figure}[!htb]
	\centering
	\includegraphics[width=1\textwidth]{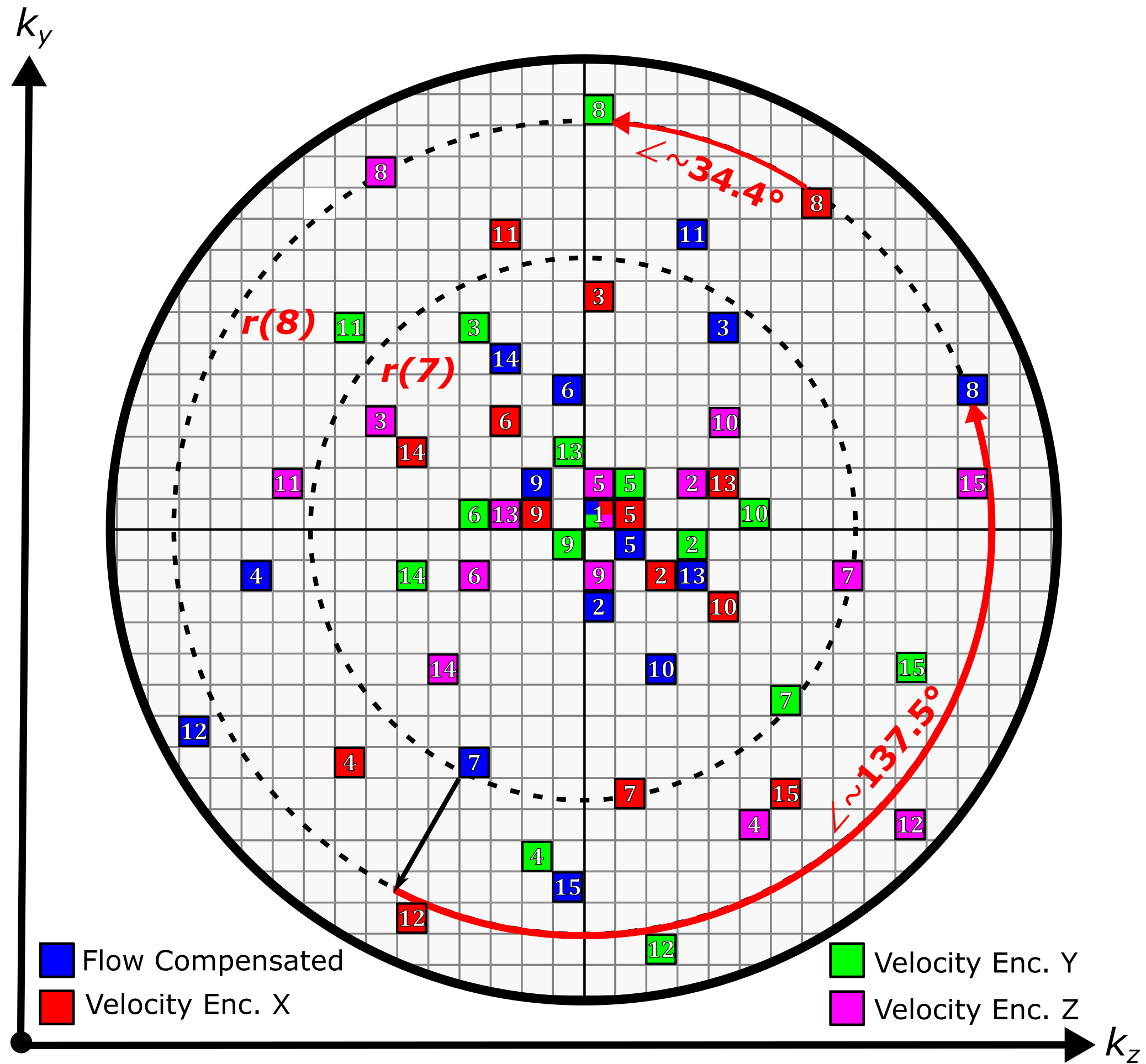}
	\caption{Illustration of the Cartesian 4D flow sampling pattern as described in the Theory section on a simplified $30\times 30$ k-space grid. The first 15 indices for each velocity encoding are shown, with initialization at the center. Self-gating indices are omitted for clarity. Each entry represents a readout along $k_x$. Dashed lines indicate radial lines for $i=7$ and $i=8$. Angular indices are advanced by $2\pi\left(2-g_\theta\right)\approx 137.5^{\circ}$, where $g_\theta$ is the golden ratio.}
	\label{fig:sampling}
\end{figure}

	Generated polar coordinates are then mapped to the $n_y \times n_z$ Cartesian grid to obtain $k_y$-$k_z$ sampling indices as
	\begin{eqnarray}
	\label{eq:ind}
	k_y(e,i) &=& { \text{round}\Big(\frac{\tilde{r}(i)}{\max(\tilde{r})}\Big\lfloor \frac{n_y}{2}-1 \Big\rfloor \sin\left(\theta(e,i)\right) + \  \Big\lfloor \frac{n_y}{2}+1 \Big\rfloor \Big)	},
	\\
	k_z(e,i) &=& { \text{round}\Big(\frac{\tilde{r}(i)}{\max(\tilde{r})}\Big\lfloor \frac{n_z}{2}-1 \Big\rfloor \cos\left(\theta (e,i)\right) + \  \Big\lfloor \frac{n_z}{2}+1 \Big\rfloor \Big)	},
	\end{eqnarray}
	where $k_y(e,i)$ and $k_z(e,i)$ represent the $i^\text{th}$ readout indices in the phase-encoding and slice-encoding directions, respectively, from the $e^\text{th}$ encoding. A representation of the described sampling is illustrated in Figure \ref{fig:sampling}. Once readout indices are generated, the self-gating lines passing through the center of k-space are injected at regular intervals, yielding the final sampling pattern.

	\subsection{Physiological Motion Signal Extraction and Binning}
	The mean signal intensity contained within a given imaging volume (e.g. the DC component) is encoded in the amplitude of the center point, $k_0$, in k-space. Physiological processes, such as the cardiac and respiratory cycles, are known to modulate the $k_0$ amplitude via changes in the blood pool volume and organ displacement, granting insight into the current physiological state~\cite{Larson2004}. Acquisition of a readout passing through $k_0$ extends this concept, providing not only the overall signal modulation, but also displacement information along a salient imaging dimension. Repetitive readouts through the center of k-space thus may provide a reliable way to track these physiological motion components.
		
	To obtain surrogate signals for cardiac and respiratory motion, the interleaved self-gating readout lines from the sampling pattern described above are extracted from the raw k-space and Fourier transformed yielding a series of complex-valued 1-D projections in the image domain. The magnitude of projections is then organized chronologically into a ($n_{RO}\times n_{C}$) $\times$ $n_{L}$ Casorati matrix, $M$, where $n_{RO}$, $n_{C}$, and $n_{L}$ correspond to readout length, number of channels, and number of acquired self-gating lines, respectively. Because spectral components of the cardiac and respiratory profiles tend to reside within relatively narrow and predictable temporal frequency bands, temporal filtering provides an effective way to mitigate noise, eliminate erroneous and non-physiological signals, and enforce separation between the physiological components. To this end, two parallel filtering operations are performed along the $j^\text{th}$ row of $M$, $M(j)$
	\begin{eqnarray}
	\label{eq:filC}
	M_C(j) &=& {	M(j)\otimes F_C},	\\
	\label{eq:filR}
	M_R(j) &=& {	M(j)\otimes F_R},
	\end{eqnarray}
	where $F_C$ and $F_R$ represent finite impulse response band-pass filters chosen with frequency cutoffs corresponding to physiological knowledge of the cardiac ($\tau_{C,low}$, $\tau_{C,high}$) and respiratory ($\tau_{R,low}$, $\tau_{R,high}$) motion signal components. The operator `$\otimes$' denotes linear convolution. The filtering process is repeated for all values of $j=1,2,...,(n_{RO}\times n_{C})$. A principal component analysis is then performed on each filtered Casorati matrix, $M_C$ and $M_R$, yielding
	\begin{eqnarray}
	\label{eq:pcaC}
	v_C &=& {	\mathcal{V}\Big(M_C\Big)},	\\
	\label{eq:pcaR}
	v_R &=& {	\mathcal{V}\Big(M_R\Big)},
	\end{eqnarray}		
	where $v_C$ and $v_R$ are vectors of size $n_L$ and represent the surrogate signals for cardiac and respiratory motion, respectively, and $\mathcal{V}(.)$ denotes extraction of the right-singular vector corresponding to the largest singular value.
		
	Once extracted, the cardiac surrogate signal undergoes a trigger detection process prior to data binning. First, zero-crossings with largest absolute slope, $|v_C(l+1)-v_C(l)|$ for $l\in 1,2,...,n_L$, are identified for each cardiac cycle. The peak that immediately precedes an identified zero-crossing is selected as a trigger point for that RR interval. If desired, $v_C$ may be initially interpolated for increased precision. Finally, detected triggers are used to segment each cardiac cycle into a user-defined number of cardiac phases and to assign each readout to a cardiac phase.
		
\begin{figure}[!htb]
	\centering
	\includegraphics[width=1\textwidth]{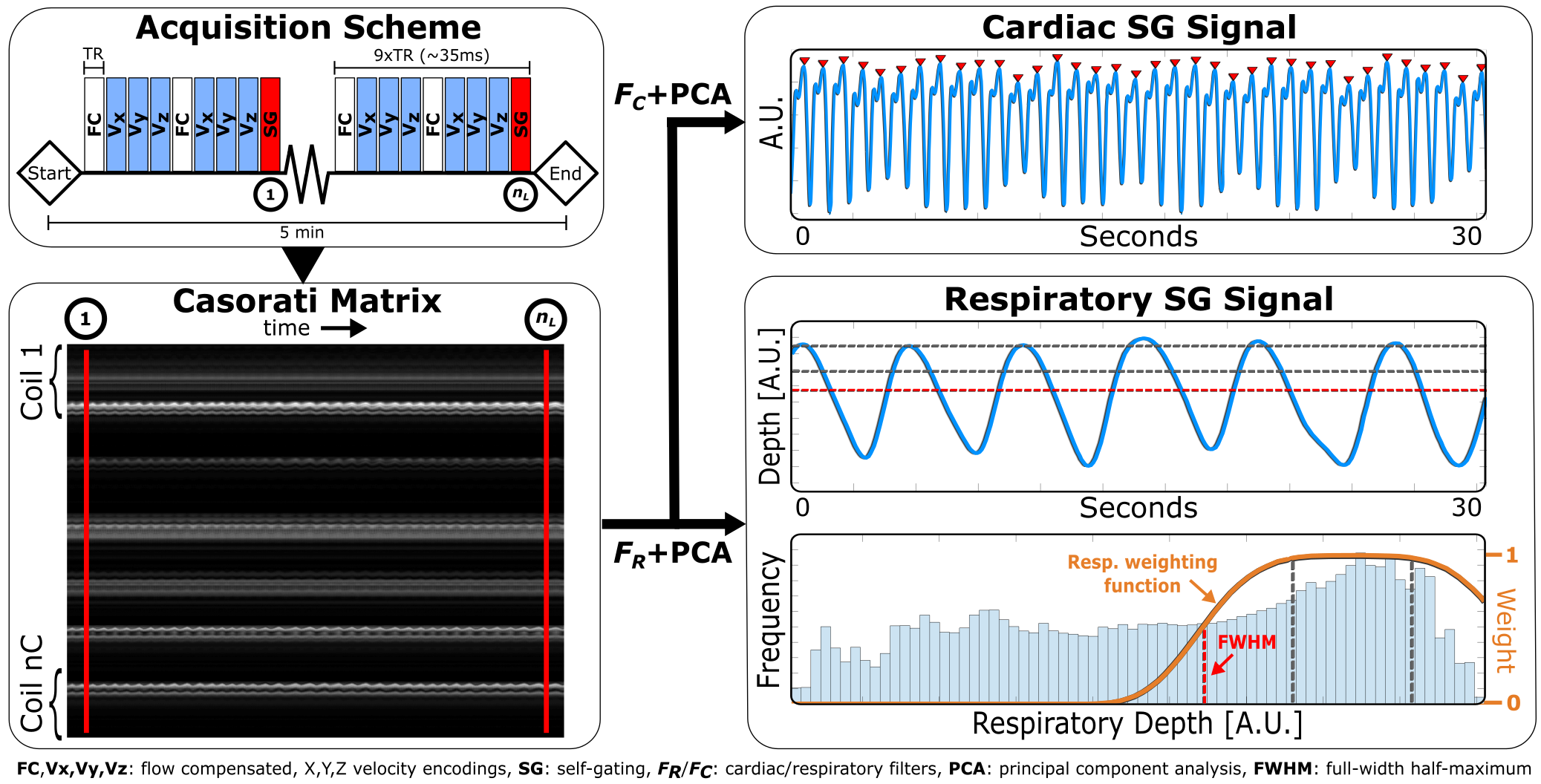}
	\caption{Self-gating (SG) signal extraction and processing pipeline. First, the interleaved self-gating readout lines through the center of the $k_y$-$k_z$ k-space plane are extracted and reorganized into Casorati matrix, $M$. Two parallel filtering operations are performed along the temporal dimension (rows) of $M$, followed by principal component analysis (PCA) to extract cardiac and respiratory motion surrogate signals. Detected peaks (red triangles) are shown on the cardiac signal and used for cardiac binning. The weighting function overlaid on the respiratory amplitude histogram is centered at end-expiration and incorporated into the reconstruction process. Parameters used for this example are as follows: $TR=3.95$ ms, SG sampling frequency$~=9\,TR$, acquisition time $=5$ min, readout direction $=$ superior-to-inferior (SI), $\tau_{C,low}$/$\tau_{C,high}$ $=0.5/3$ Hz,  $\tau_{R,low}$/$\tau_{R,high}$ $=0/0.5$ Hz, and respiratory efficiency $=50\%$. Only 30 seconds of the motion signals are shown for brevity. }
	\label{fig:self_gating}
\end{figure}

	To compensate for respiratory motion, a generalized soft-gating approach is used~\cite{Cheng2013}. The relative amplitude of the respiratory surrogate signal, $v_R$, is used to define a weighting function that penalizes the relative contributions of measured k-space lines during the reconstruction process. The respiratory weighting function with center, $\mu$, and order, $p\geq1$, may be defined for the $i^\text{th}$ readout as
	\begin{equation}
	\label{eq:weight}
	W(i) = {\exp\left\{-\frac{(\tilde{v}_R(i)-\mu)^p}{\phi}\right\}},
	\end{equation}		 
	where $\tilde{v}_R$ represents an interpolated version of $v_R$ such that $\tilde{v}_R(i)$ is defined for every readout and not just the self-gating readouts. The center, $\mu$, selects the desired respiratory state. The order, $p$, determines the shape of function, $W$. The width of $W$ is controlled by the parameter, $\phi$; the value for $\phi$ may be automatically calculated to achieve a desired respiratory efficiency, where efficiency, in percent, is computed as $100\times \sum_{i=1}^{n_S}\frac{W^2(i)}{n_S}$, where $n_S$ is the total number of acquired k-space readouts. Alternatively, $\phi$ can be chosen to achieve a specific acceleration rate, computed as $\sum_{i=1}^{n_S}\frac{n_M}{W^2(i)}$, where $n_M$ represents the number of elements in the imaging matrix. If reconstructing $m>1$ respiratory phases is desired, multiple weighting functions may be defined with different values of $\mu_1,\cdots,\mu_m$ and $\phi_1,\cdots,\phi_m$ for a given dataset. The self-gating signal extraction and processing pipeline described in this section is summarized in Figure \ref{fig:self_gating}.

\section{METHODS}
	\subsection{MR Acquisition - Healthy Subjects}	
	The proposed 4D flow imaging pipeline was validated in eight healthy subjects (age range, 20--46 years; two female) prospectively recruited for this study. Each subject provided informed, written consent in accordance with the local ethics board prior to scanning. All subjects were imaged using a clinical 3T MR system (MAGNETOM Prisma, Siemens Healthcare, Erlangen, Germany) equipped with a 18-channel chest coil and spinal array. %All exams were completed within one hour.
			
		\subsubsection{Whole-heart 4D flow imaging}
		The prototype 4D flow imaging sequence was based on a spoiled gradient echo readout with an asymmetric four-point encoding scheme and was acquired free-running and free-breathing for a fixed acquisition time of five-minutes; sampling, as described in the Theory section, was generated inline with self-gating lines acquired every nine readouts. The imaging volume was prescribed as a sagittal slab covering the whole-heart and aortic arch with fixed matrix size of $96\times96 \times 56$ (readout $\times$ phase encoding $\times$ slice encoding); readout was  oriented along the superior-inferior (SI) direction. Spatial resolution was subject-dependent but kept isotropic between $2.3$ and $2.6$ mm. %Other relevant acquisition parameters were: TE $=2.17$~ms, TR $=3.95$~ms, flip angle $=7^{\circ}$, VENC $=150$~cm/s, BW $=801$~Hz/px, FOV $=(220-250)\times (220-250)\times (129-146$)~mm, and slice oversampling $=28$\%. 
		Other relevant acquisition parameters are summarized in Table \ref{tab:scan_param}.
				
		\subsubsection{2D phase contrast}
\begin{figure}[!htb]
	\centering
	\includegraphics[width =1 \textwidth]{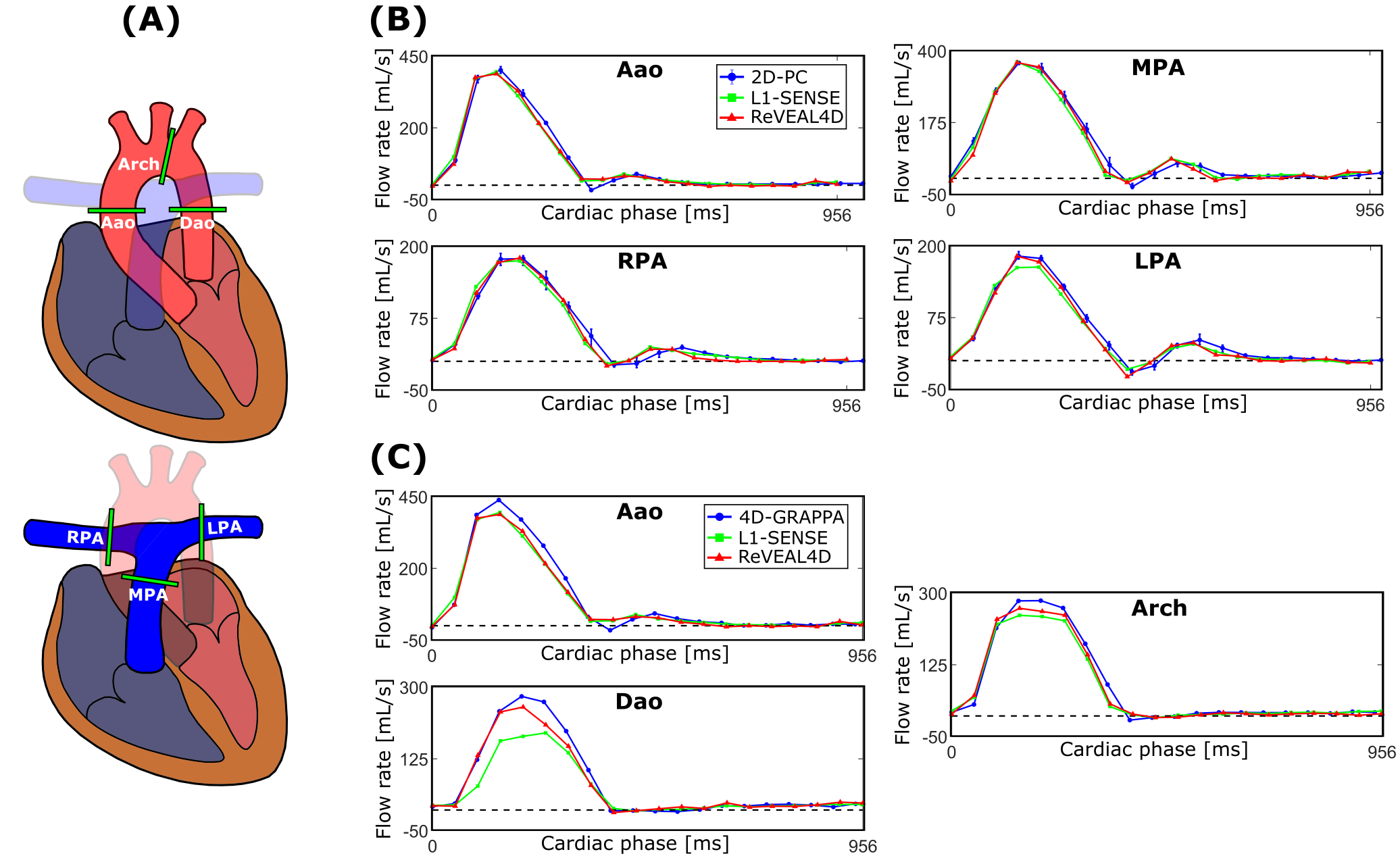}
	\caption{(A) Analysis planes defined for 4D flow and 2D phase contrast over which flow quantification is performed. Representative volumetric flow rate profiles measured from the reconstructed whole-heart 4D flow images compared with (B) 2D phase contrast (2D-PC) and (C) GRAPPA accelerated and navigator-gated 4D flow called 4D-GRAPPA. Error bars corresponding to 2D-PC represent the range over three repeats. L1-SENSE and ReVEAL4D reconstructions share identical binned k-space and segmentation.}
	\label{fig:segmentation_and_flow_curves}
\end{figure}

		As reference for measured flow indices, breath-held (expiration), segmented 2D phase-contrast (2D-PC) images were acquired in all subjects using typical clinical protocols, with retrospective ECG gating (20 cardiac phases) and GRAPPA ($R=2$) reconstruction. Slice planes were prescribed transecting the ascending aorta (Aao), main pulmonary artery (MPA), right pulmonary artery (RPA), and left pulmonary artery (LPA) (Figure \ref{fig:segmentation_and_flow_curves}). Each vessel was imaged a total of three times over the course of the exam to account for natural physiological flow variability. %All images were reconstructed on the scanner to generate 20 cardiac phases; 
		Additional acquisition parameters are summarized in Table \ref{tab:scan_param}.

		\subsubsection{Navigator-gated 4D flow imaging}
		A secondary reference, herein referred to as 4D-GRAPPA, in the form of a vendor-provided, segmented, 4D flow imaging prototype sequence was also acquired in all subjects. Images were acquired during free-breathing with respiratory navigator gating (acceptance window $\pm 5$ mm),  retrospective gating (20 cardiac phases), and accelerated by GRAPPA ($R=3$). The imaging volume was prescribed as a single-oblique (sagittal to coronal) slab covering the entire aorta. %Images were reconstructed on the scanner to generate 20 cardiac phases. 
		Additional scan parameters are listed in Table \ref{tab:scan_param}.
		
    \subsection{MR Acquisition - Patient\textk{s}}
        \textk{The data using the proposed 4D flow sequence were acquired as an addendum to the clinically prescribed CMR exam in two patients. The first patient (Patient 1) was a 37-year-old male with a history of congenital pulmonary stenosis and right pulmonary artery stenosis post Blalock-Taussig shunt and valvatomy. The second patient (Patient 2) imaged was an 80-year-old male with hypertension, coronary artery disease, and a high burden of premature ventricular contractions. Patients provided informed, written consent and the study was conducted in accordance with the local ethics board and HIPAA regulations. Imaging was carried out on a 1.5T clinical scanner (MAGNETOM Sola, Siemens Healthcare, Erlangen, Germany) equipped with a 12-channel cardiac receiver array and spine coil.}
        \subsubsection{Whole-heart 4D flow imaging}
        \textk{The prototype self-gated 4D flow imaging sequence previously described for the 3T MR system was adapted to the 1.5T system for acquisition in patients. The sampling pattern, including self-gating readouts every nine lines, was automatically generated on the scanner. Acquisitions were performed free-running and free-breathing for a total of five minutes with an imaging volume prescribed as a sagittal slab covering the whole-heart and great vessels. Readouts were oriented along the SI direction. 
        Acquisition parameters for 4D flow imaging are summarized in Table \ref{tab:patient_parameters}. Alongside 4D flow acquisition, synchronous ECG signals were also recorded.}
        
        \subsubsection{2D phase contrast}
        \textk{Breath-held 2D-PC images were acquired in both patients as a component of their CMR examination using typical clinical protocols, retrospective ECG triggering (20 cardiac phases), and rate-2 acceleration. In Patient 1, slice planes were prescribed transecting the Aao, MPA, RPA, and LPA; in Patient 2, only the Aao was imaged. %Images were reconstructed on the scanner to generate 20 cardiac phases. 
        Acquisition parameters for 2D-PC are summarized in Table \ref{tab:patient_parameters}.}		
		
	\subsection{Binning and Image Recovery}
	Once the whole-heart 4D data were acquired, cardiac and respiratory motion surrogate signals, $v_C$ and $v_R$, were extracted from the self-gating lines as detailed in the Theory section. For $F_C$ and $F_R$, frequency cutoffs of [0.5, 3] Hz and [0, 0.5] Hz, respectively, were chosen. Data were binned into 20 cardiac phases, yielding a temporal resolution from 43 ms to 65 ms \textk{(R-R intervals from 860 ms to 1300 ms)} across all \textk{healthy} subjects \textk{and 45 ms to 65 ms (R-R intervals from 904 ms to 1302 ms) in patients}. Bin widths were fixed per subject and determined from mean \textk{R-R interval}. Beginning from the first detected cardiac trigger, bins were propagated forward through the cardiac cycle in a manner analogous to prospective triggering until (i) the bin edge reached the following trigger or (ii) all 20 phases had been defined for the given cycle. This process was repeated for all detected triggers. Arrhythmic or outlier heart beats, defined as greater than three standard deviations away from the mean \textk{R-R interval}, were discarded from consideration.
	
	One weighting function, $W$, corresponding to end-expiration was defined for each subject. A two-compartment Gaussian mixture model was fit to the $\tilde{v}_R$ histogram; the center, $\mu$, of the respiratory weighting function was determined as the mean of the most prominent component. This assumes that the most populated part of the histogram belongs to end-expiration. The exponential order, $p$, was set to 4, yielding a weighting function similar to the one shown in Figure \ref{fig:self_gating}.  Width of the weighting function, $\phi$, was automatically chosen such that respiratory efficiency (see Theory) was fixed at $50\%$ for all subjects. \textk{In the event of duplicate binned k-space samples, the sample corresponding to the maximum respiratory weight, $W$, was selected; remaining duplicates were discarded.}
	
	The binned and weighted k-space data were reconstructed using the recently proposed ReVEAL4D algorithm \cite{Rich2016,Rich2019}. ReVEAL4D exploits spatio-temporal sparsity in the wavelet domain as a means of regularization. Additionally, ReVEAL4D explicitly enforces magnitude and phase similarities between velocity encodings, while preserving phase differences encoded within regions of flow so as to jointly reconstruct the flow compensated and flow encoded images.
	ReVEAL4D iteratively computes the minimum mean squared estimate for the Bayesian signal model using the GAMP message passing algorithm~\cite{Rangan2011}. For comparison, all binned k-space were additionally reconstructed using the conventional L1-SENSE CS framework~\cite{Liu2008}, whereby enforcing only spatio-temporal sparsity in the wavelet domain was used as a regularizer. The implementation of L1-SENSE was based on the bFISTA solver~\cite{ting2017fast}. The tuning parameters for ReVEAL4D and L1-SENSE were empirically chosen based on performance in an additional dataset not included in this study. \textk{Tunable parameters for ReVEAL4D included wavelet regularization strength, $\lambda$, extent of inter-encoding mismatch due to departure from ideal physics, $\sigma^{2}$, and voxel-wise prior probability for the presence of flow, $\gamma$~\cite{Rich2019}. The values of these parameters were manually optimized using two additional datasets not included in this study and set at $\lambda=1.5$, $\sigma^2=0.01$, and $\gamma=0.95$.
	%, in healthy subjects. In the patient cohort, $\lambda$ was set to $0.3$ while $\sigma^{2}$ and $\gamma$ remained constant.These parameters are dependant on both noise power and data scaling. 
	One parameter, $\lambda$, was tuned for L1-SENSE and set at $5 \times 10^{-4}$.}
	
	In both reconstructions, coil sensitivities were estimated from time-averaged k-space using a method by Walsh et al. \cite{Walsh2000}. Reconstructions were implemented offline in custom MATLAB software (Mathworks, Natick, MA) utilizing combined GPU/CPU computation on a Windows 10 workstation equipped with an Nvidia RTX 2080 Ti GPU and 10-core Intel Xeon W-2155 CPU. Typical reconstruction times for ReVEAL4D were approximately 2.5 hours per subject. Acceleration rates for the whole-heart 4D flow imaging datasets ranged from approximately 19 to 20 \textk{in healthy subjects and 23 to 26 in patients}, which did not include the additional undersampling due to the asymmetric echo.
	
	\subsection{Post-processing and Analysis}
 	 Reconstructed whole-heart 4D flow images were converted to DICOM and flow quantification was performed in Siemens 4D Flow v2.4 prototype software (Siemens Healthcare, Erlangen, Germany). \textk{For the healthy subject study,} six analysis planes were defined transecting the Aao, MPA, RPA, and LPA, as well the aortic arch (Arch) and descending aorta (Dao) (Figure \ref{fig:segmentation_and_flow_curves}). Planes transecting the Aao, MPA, RPA, and LPA were manually selected to best match corresponding planes acquired with 2D-PC. The contours were semi-automatically drawn for each vessel for flow quantification. Identical analysis planes and contours were used for both the ReVEAL4D and L1-SENSE reconstructions. The conventional aortic 4D flow images were similarly segmented along the Aao, Arch, and Dao, using the same analysis planes defined for the whole-heart images. The 2D-PC images (three repeats of Aao, MPA, RPA, and LPA, per subject) were semi-automatically segmented and analyzed using the freely-available software, Segment v2.2 R7056 (http://segment.heiberg.se) \cite{Heiberg2010}. Velocity and pathline visualizations were generated using the cvi42 4D flow module (Circle Cardiovascular Imaging, Calgary, Canada). \textk{Segmentation of the whole-heart 4D flow and 2D-PC images was analogously performed for the patient study, wherein the Aao, MPA, RPA, and LPA were segmented in the first patient, and the Aao segmented in the second. Patient images were interpolated to 2.5 mm isotropic spatial resolution prior to segmentation and analysis. }
 	
 	Prior to flow quantification, all images were corrected using a recently proposed background phase correction method ~\cite{Pruitt2019}. Phase-aliasing correction was performed when necessary. Net volumetric flow, peak through-plane velocity, and peak volumetric flow rate were calculated as \textk{the flow indices of interest}. Peak velocities were calculated following local $3\times 3\times 3$ median filtering to guard against erroneous inclusion of noise pixels. 
 	
 	\textk{In healthy subjects,} indices derived from 2D-PC were taken as the average of the three repetitions. Correlation statistics of the flow indices derived from the whole-heart 4D flow images were calculated with respect to 2D-PC and conventional 4D flow imaging. Bland-Altman analysis was additionally performed, treating 2D-PC and conventional 4D flow imaging as separate references; bias and limits of agreement (LOA) were reported in terms of percent error. Biases were reported as significant for \textit{P}-value, {$\textit{P}<0.05$}, via independently performed paired t-tests. Reported statistics represent an aggregate of quantified vessels. \textk{In patients, net flow, peak through-plane velocity, and peak flow rate were quantified in the analyzed vessels. Indices derived from whole-heart 4D flow and 2D-PC were reported for each patient and vessel. Volumetric flow rate curves encompassing the complete cardiac cycle were also reported.} 
 	
 	\textk{Additionally, cardiac SG signals were compared with the \textk{ECG signals recorded from both patients during acquisition.}  Mean R-R interval was calculated from accepted triggers for the SG and ECG signals over the five-minute scan duration. Trigger precision error of the SG signal with respect to ECG was also evaluated according to the following formula:
 	\begin{equation}
 	\label{eq:trigger precision}
 	    \text{\emph{Precision Error}} = stdev(TT_{SG}(t) - TT_{ECG}(t)),
 	\end{equation}
 	where $TT_{SG}$ and $TT_{ECG}$ represent trigger times from SG and ECG, respectively, $stdev(*)$ is the standard deviation, and $t$ indicates the cardiac cycle index of accepted SG triggers with a corresponding ECG trigger present within a $\pm200$ ms search window centered at the SG trigger time. }

\section{RESULTS}
\subsection{Healthy subjects}
\begin{figure}[!htb]
	\centering
	\includegraphics[width =1\textwidth]{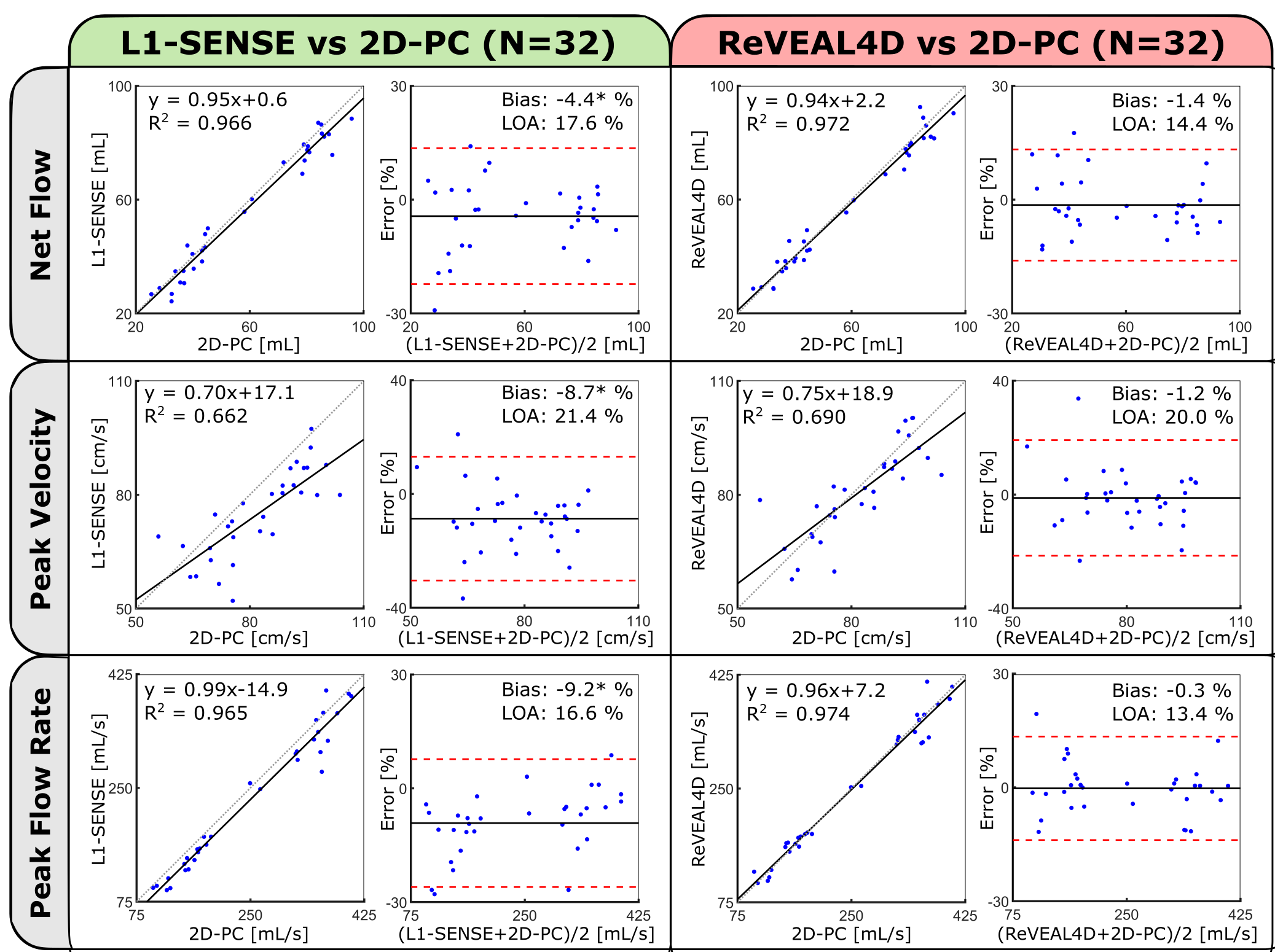}
	\caption{Comparison of measured flow metrics (net flow, peak velocity, and peak flow rate) calculated from the reconstructed whole-heart 4D flow and 2D phase contrast (2D-PC) using correlation and Bland-Altman analysis. Statistics are derived from the aggregate of Aao, MPA, RPA, and LPA measurements in eight subjects. Individual 2D-PC measurements represent the mean of three repeat acquisitions. Bland-Altman statistics are reported in terms of percentage error. Significant biases ($P<0.05$) are indicated by asterisks. }
	\label{fig:BA_2D}
\end{figure}

Figure \ref{fig:BA_2D} shows the comparison between net flow (top row), peak through-plane velocity (middle row), and peak flow rate (bottom row) quantified from the reconstructed whole-heart 4D flow imaging and 2D-PC images, aggregated for the Aao, MPA, RPA, and LPA ($N=32$). The ReVEAL4D reconstruction demonstrated strong agreement with respect to the 2D-PC reference for all indices. Bland-Altman analysis statistics (bias$\pm$LOA, in $\%$ error) give $-1.4\pm 14.4\%$ for net flow, $-1.2\pm 20.0\%$ for peak through-plane velocity, and $-0.3\pm 13.4\%$ for peak flow, with non-significant biases relative to 2D-PC. For L1-SENSE reconstruction, the underestimations were more pronounced, yielding $-4.4\pm 17.6\%$ for net flow, $-8.7\pm 21.4\%$ for peak through-plane velocity, and $-9.2\pm 16.6\%$ for peak flow. The resulting biases relative to 2D-PC were significant for peak through-plane velocity ($P<0.0001)$ and peak flow ($P<0.0001)$. These results are additionally corroborated by the given correlation plots.
	
\begin{figure}[!htb]
	\centering
	\includegraphics[width = 1\textwidth]{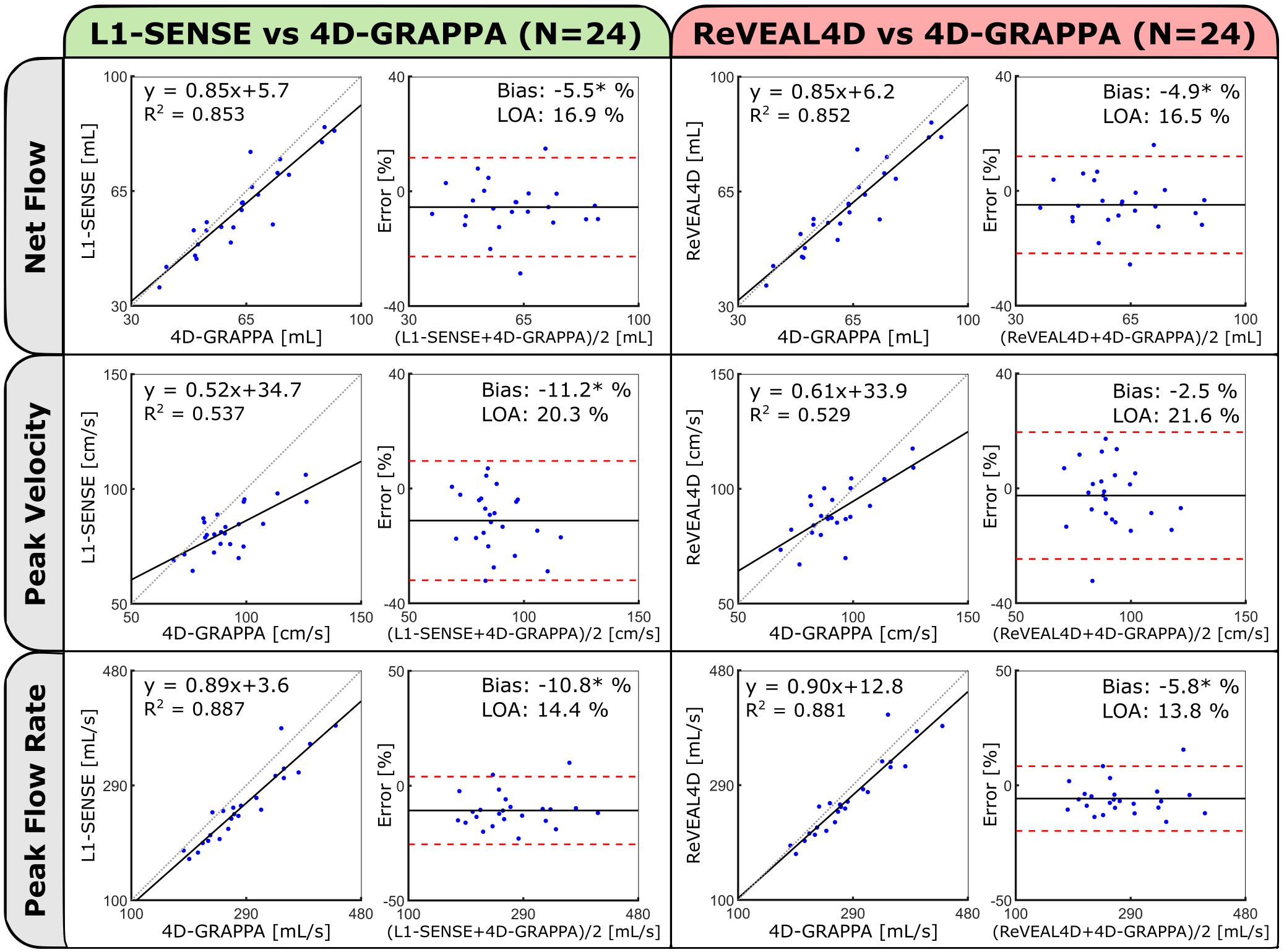}
	\caption{Comparison of measured flow metrics (net flow, peak velocity, and peak flow rate) calculated from the reconstructed self-gated, whole-heart 4D flow and conventional 4D flow using correlation and Bland-Altman analysis. Statistics are derived from the aggregate of ascending aorta, aortic arch, and descending aorta measurements in eight subjects. Bland-Altman statistics are reported in terms of percentage error. Significant biases ($P<0.05$) are indicated by asterisks.}
	\label{fig:BA_4D}
\end{figure}

Likewise, Figure \ref{fig:BA_4D} compares the measured flow indices quantified from the proposed whole-heart 4D flow image reconstructions and conventional 4D-GRAPPA in the aggregated  Aao, Arch, and Dao ($N=24$). For ReVEAL4D reconstruction, Bland-Altman analysis statistics are: $-4.9\pm 16.5\%$ for net flow, $-2.5\pm 21.6\%$ for peak through-plane velocity, and $-5.8\pm 13.8\%$ for peak flow, with small but significant biases for net flow ($P=0.0106$) and peak flow ($P=0.0006$). For L1-SENSE reconstruction, the underestimations were more pronounced, yielding $-5.5\pm 16.9\%$ for net flow, $-11.2\pm 20.3\%$ for peak through-plane velocity, and $-10.8\pm 14.4\%$ for peak flow. The resulting biases relative to 4D-GRAPPA were significant for net flow ($P=0.0051$), peak through-plane velocity ($P<0.0001)$ and peak flow ($P<0.0001)$. Similarly, these results are also corroborated by the presented correlation plots.

\begin{figure}[!htb]
	\centering
	\includegraphics[width=1\textwidth]{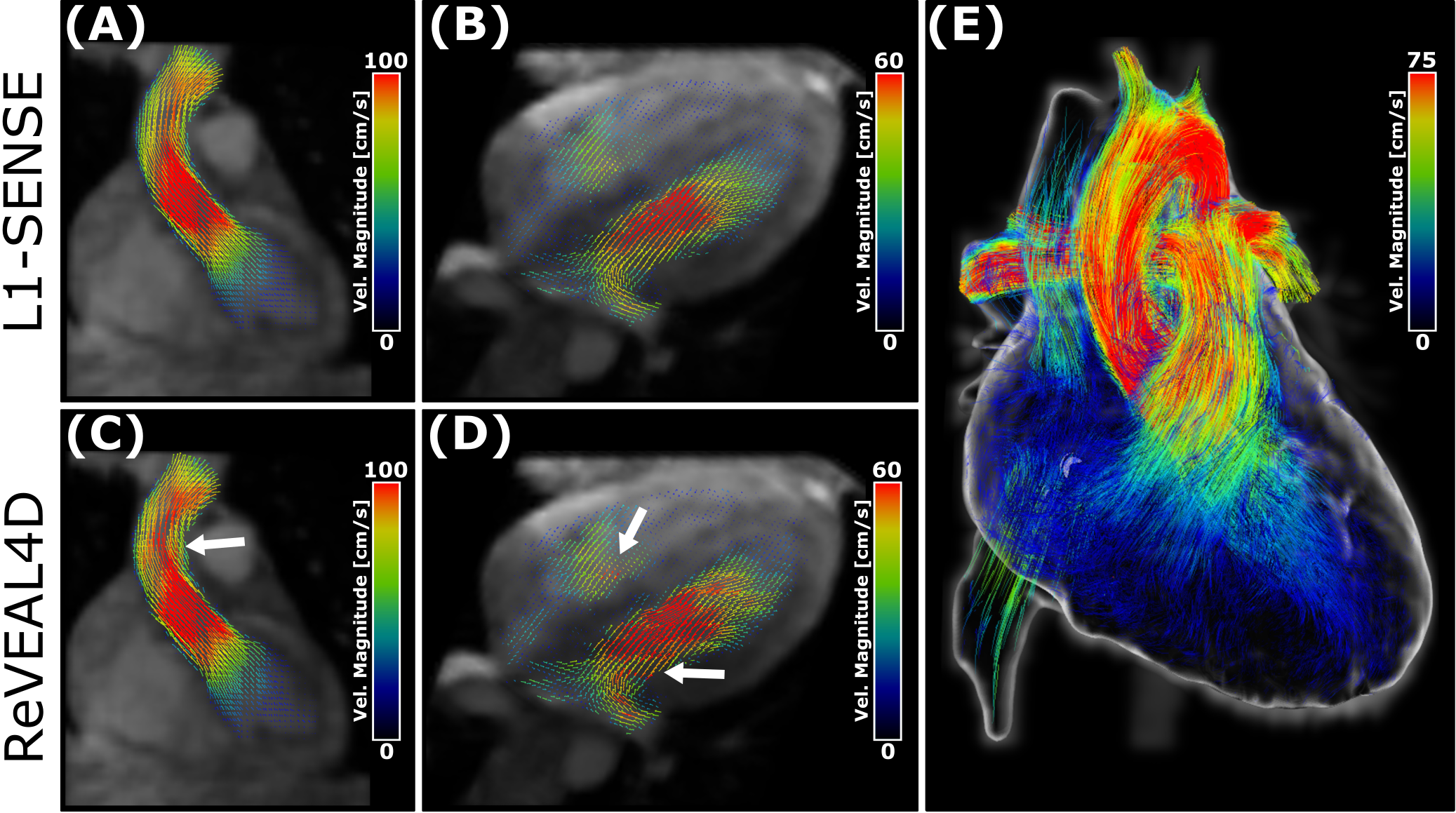}
	\caption{\textk{Reconstructed 4D flow images with velocity vector overlays from a representative subject. Images reconstructed with L1-SENSE (top row) and ReVEAL4D (bottom row). Displayed images are reformatted to (A,C) left-ventricular outflow tract, (B,D) four-chamber view, and (E) whole-heart velocity pathline rendering of the same subject at systole using ReVEAL4D reconstruction. White arrows highlight notable velocity discrepancies between the L1-SENSE and ReVEAL4D reconstructions in the (C) ascending aorta and (D) mitral and tricuspid inflows. For each view, the cardiac phase with most the salient flow features is shown.}}
	\label{fig:reconstructed_images}
\end{figure}

\begin{figure}[!htb]
	\centering
	\includegraphics[width =.75\textwidth]{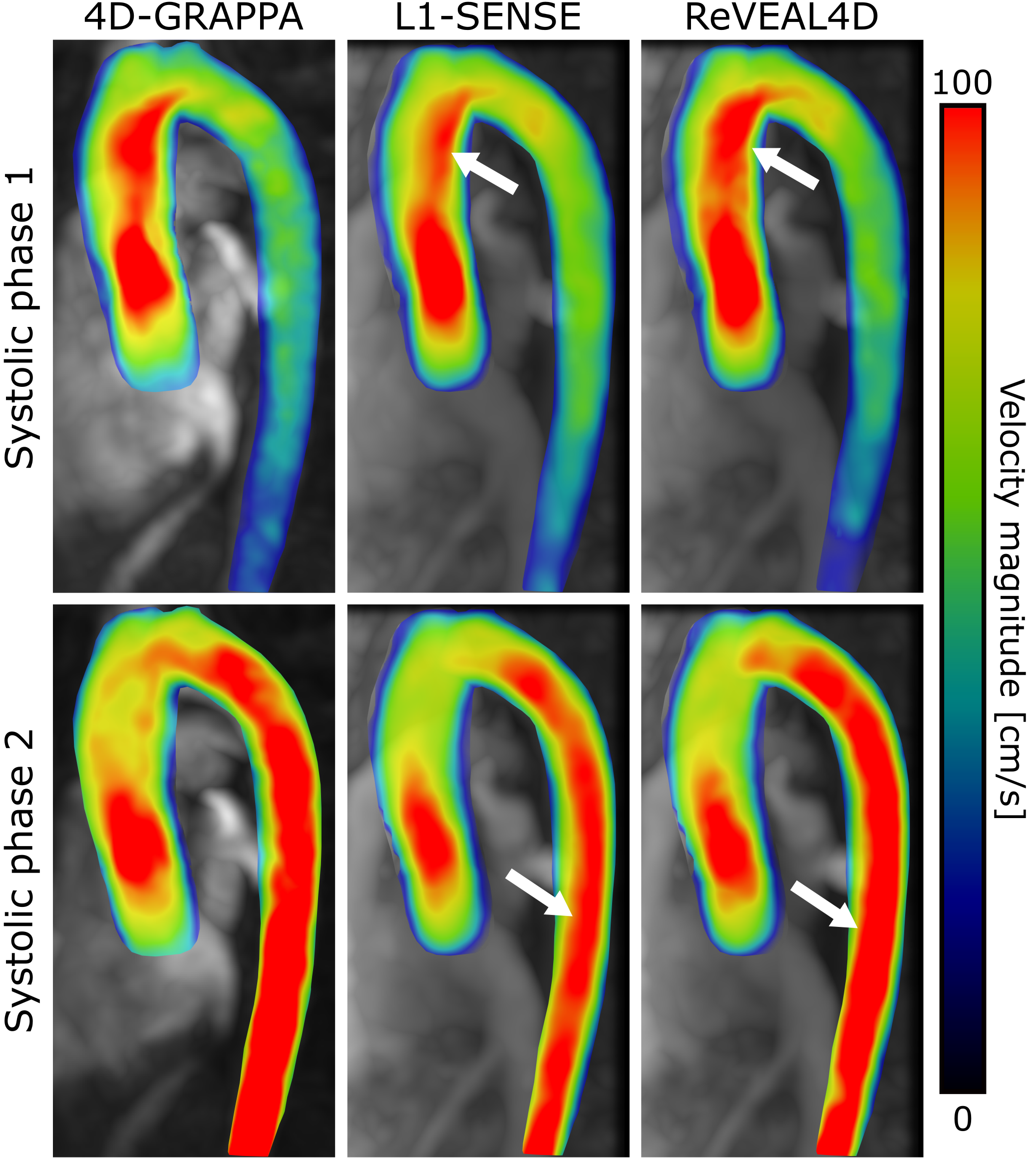}
	\caption{Aorta segmented from conventional and self-gated, whole-heart 4D flow images with velocity magnitude overlay. The top and bottom rows depict different systolic frames with flow predominantly in the ascending aorta and descending aorta, respectively. Arrows indicate regions of depressed velocity in the L1-SENSE reconstructed image compared with ReVEAL4D and 4D-GRAPPA.}
	\label{fig:MIP_aorta}
\end{figure}

\textk{Representative magnitude images superimposed with velocity vectors from the left ventricular outflow tract (LVOT) and four-chamber view (4ch)} are shown in Figure \ref{fig:reconstructed_images}\textk{A-D} at individually salient cardiac phases. Compared with L1-SENSE reconstruction, ReVEAL4D provided noticeably increased velocity estimation in the LVOT and 4ch views (white arrows), in agreement with the results presented in Figures \ref{fig:BA_2D} and \ref{fig:BA_4D}. Whole-heart pathline visualization at a systolic frame is shown in Figure \ref{fig:reconstructed_images}E. Figure \ref{fig:MIP_aorta} shows velocity maximum intensity projections of the aorta, segmented from the conventional 4D flow imaging, L1-SENSE, and ReVEAL4D images at two different systolic frames corresponding to peak flow rate in the Aao (top row) and Dao (bottom row). White arrows highlight clear velocity discrepancies  between L1-SENSE and ReVEAL4D reconstructions, whereby ReVEAL4D results in more similar velocity estimation with respect to the conventional 4D flow imaging reference, 4D-GRAPPA. 
\subsection{Patient\textk{s}}
\begin{figure}[!htb]
	\centering
	\includegraphics[width =0.65\textwidth]{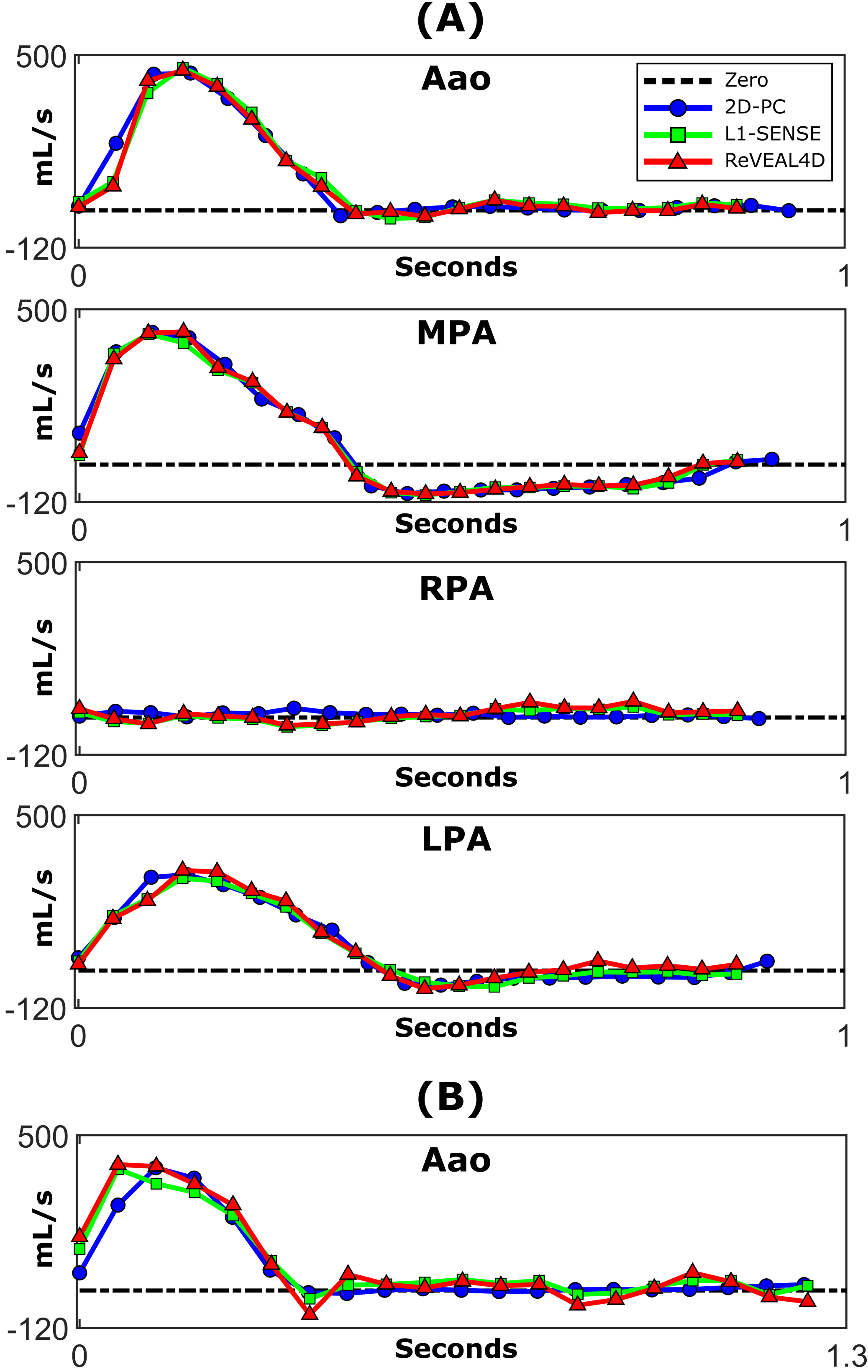}
	\caption{\textk{Volumetric flow rate curves analyzed from (A) Patient 1 and (B) Patient 2. Volumetric flow rates are displayed for 2D-PC and the two reconstruction methods used for whole-heart 4D flow (L1-SENSE and ReVEAL4D). The low flow observed in the RPA of Patient 1 was due to severe stenosis in this vessel. }} 
	\label{fig:patient_flow}
\end{figure}

\textk{Figure \ref{fig:patient_flow} depicts volumetric flow rate curves for each vessel analyzed in \textk{both patients.} Flow rates are shown for whole-heart 4D flow imaging (L1-SENSE and ReVEAL4D) and compared with conventional 2D-PC. In Patient 1 (Figure \ref{fig:patient_flow}A), L1-SENSE and ReVEAL4D both demonstrate good agreement with respect 2D-PC in the Aao, MPA, RPA, and LPA. Similar agreement between L1-SENSE/ReVEAL4D and 2D-PC may be observed in the Aao analyzed from Patient 2 (Figure \ref{fig:patient_flow}B). Table \ref{tab:patient_flow} summarizes the flow indices, net flow, peak through-plane velocity, and peak flow rate quantified from whole-heart 4D flow imaging and the  2D-PC reference. Peak flow rate represents the flow index with greatest mean absolute error between ReVEAL4D and 2D-PC, with a mean error of 10.8 mL/s and range from -0.8 mL/s to 22 mL/s \textk{across the two patients.} Similarly, L1-SENSE exhibited the greatest error in peak through-plane velocity estimation, with a mean error of -19.1 cm/s ranging from -30.7 cm/s to -8.1 cm/s. Mean errors for the remaining flow indices are reported as follows: net flow and peak velocity derived from ReVEAL4D are -0.5 mL and -3.8 cm/s, respectively; net flow and peak flow rate derived from L1-SENSE are -2.6 mL and -0.3 mL/s, respectively.}

\begin{figure}[h]
	\centering
	\includegraphics[width =1 \textwidth]{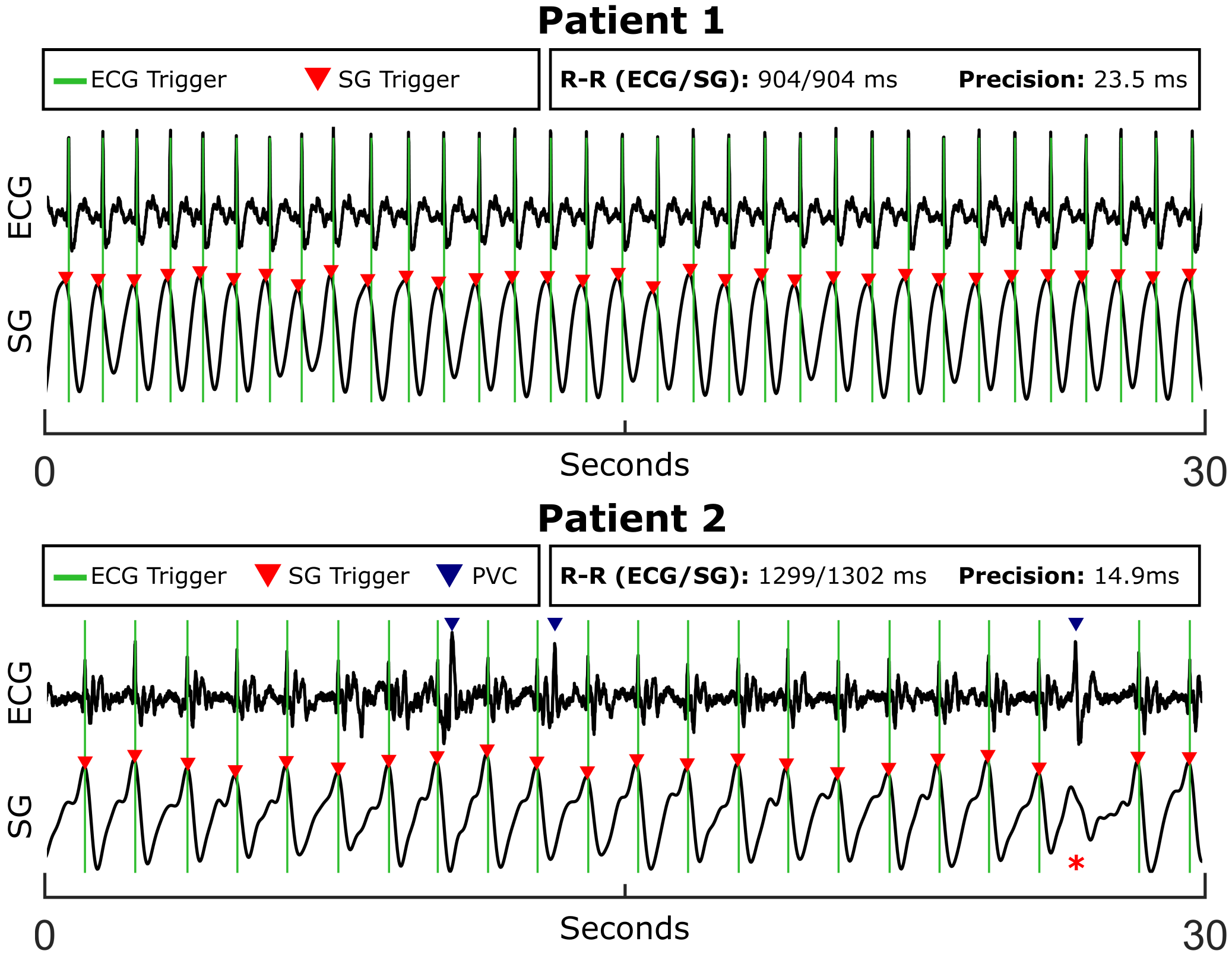}
	\caption{\textk{Comparison between the derived SG signal and synchronous ECG acquired in \textk{patients.} Trigger points derived from ECG and SG are marked on the signal traces. Note the presence of premature ventricular contraction (PVC) in Patient 2 (blue triangles) as visible in the ECG.  An example of an automatically rejected SG trigger coinciding with a PVC is denoted by the asterisk. However, not all PVC triggers manifested in the SG signal, which captures the mechanical motion of the heart.}} 
	\label{fig:patient_ecg}
\end{figure}

\textk{Comparisons between synchronously acquired ECG and SG triggers along with signal traces covering a 30-second period are given for each patient in Figure \ref{fig:patient_ecg}. In Patient 1, triggers from ECG and SG result in the same R-R interval of 904 ms; precision error of the SG triggers with respect to ECG as computed from Equation \ref{eq:trigger precision} was found to be 23.5 ms. Likewise in Patient 2, R-R intervals are 1299 ms and 1302 ms from ECG and SG, respectively with a precision error of 14.9 ms. Note the high burden of premature ventricular contractions (PVCs) in Patient 2 (blue triangles); an example of a PVC successfully eliminated by the previously described arrhythmia rejection criterion is marked by an asterisk.} 

\section{DISCUSSION}
This study addresses several technical challenges that have historically contributed to the lack of clinical adoption of 4D flow imaging. Our comprehensive framework combines key innovations for 4D flow imaging: (i) a flexible Cartesian sampling with on-scanner implementation, (ii) full cardiac and respiratory self-gating, and (iii) ReVEAL4D reconstruction to enable a fixed five-minute whole-heart 4D flow examination with minimal setup overhead.  

Compared to other recent methods, the proposed sampling method offers several advantages. First, the method does not require solving an optimization problem~\cite{Han2017} and can be implemented using a few lines of code, enabling instantaneous, inline computation of the sampling pattern. Second, for the proposed sampling method, the ACS region grows progressively with data acquisition, and the sampling density is parametrically controlled, avoiding the need to explicitly specify the size of the ACS region, the number and size of concentric rings, or acceleration rate \cite{cheng2015free}. Third, since the proposed method avoids sampling along radially incrementing leaflets, it provides a more uniform coverage of the k-space over a given time window. \textk{One limitation of the proposed sampling method is the reliance on several user-defined parameters controlling, for instance, angular or radial incrementation and degree of variable density. In this work, parameters were empirically selected based on apparent uniformity of the distribution, but these parameters can be further optimized.}

Prior work has demonstrated that phase-contrast and 4D flow imaging can be susceptible to flow and velocity underestimation at high acceleration rates. Bollache et al. reported 10\% and 13\% underestimation in the quantification of flow and peak velocity, respectively, using a k-t acceleration method, while Ma et al.\ reported up to a 13\% underestimation in peak velocity and peak flow rate using a CS method \cite{Ma2019}. In contrast, the proposed framework produces excellent agreement in net volumetric flow, peak velocity, and peak flow rate with respect to our references. Flow and velocity underestimation using ReVEAL4D is non-significant or minor at high acceleration, with ReVEAL4D outperforming conventional CS (L1-SENSE) overall. The superior performance of ReVEAL4D can be attributed to explicitly modeling the expected magnitude-phase relationship between velocity encodings~\cite{Rich2016}. Specifically, ReVEAL4D enforces voxel-wise magnitude and phase similarity between encodings in stationary tissue and only magnitude similarity in regions of flow, leaving phase unconstrained. These soft constraints are incorporated along with spatio-temporal wavelet sparsity as a Bayesian prior. Together, these data suggest the proposed framework may provide a rapid, free-running approach to 4D flow imaging without significantly sacrificing quantification accuracy. Although one respiratory phase was reconstructed in this work, the soft-gating function in Eq.~\ref{eq:weight} readily generalizes for respiratory-resolved 5D flow imaging \cite{Walheim2019}, at the cost of increased computational complexity.

Flow quantification performance was \textk{first} investigated in eight healthy subjects and compared with 2D-PC and conventional 4D flow imaging. As shown in Figures \ref{fig:BA_2D} and \ref{fig:BA_4D}, both reconstruction methods yielded comparable LOA with respect to 2D-PC and 4D-GRAPPA; collectively, LOA were within 17\%, 22\%, and 16\% for net flow, peak velocity, and peak flow rate, respectively. While ReVEAL4D reconstruction showed non-significant bias in flow metrics with respect to 2D-PC, small, significant underestimations were observed with respect to 4D-GRAPPA in net flow and peak flow rate. The biases, however, were smaller compared to those of L1-SENSE. This discrepancy between the 2D-PC and 4D-GRAPPA references could potentially be attributable to several factors including natural physiological variation, the use of multiple averages for 2D-PC measurements, slightly higher temporal resolution for 4D-GRAPPA, and residual respiratory motion artifacts owing to the $\pm5$ mm navigator acceptance window. \textk{Moreover, lengthening the scan duration may reduce these biases owing to the reduced acceleration factor; however, this comes at the cost of longer examinations and increases the risk of image corruption from bulk patient motion.} Our data also suggest that there may be subtle differences in flow patterns, as shown in Figures \ref{fig:reconstructed_images} and \ref{fig:MIP_aorta}, between L1-SENSE and ReVEAL4D. These flow differences may have a tangible impact on the calculation of advanced hemodynamic parameters such wall shear stress or pressure gradient, although these were not considered as part of this work.

\textk{Flow quantification was additionally evaluated in two cardiac patients and compared with 2D-PC. Both reconstructions compared favorably with 2D-PC with the notable exception of a marked  underestimation of peak through-plane velocity derived from L1-SENSE, resulting in an average 19.1 cm/s underestimation. Analysis between ECG and SG triggers yielded comparable R-R intervals with trigger precision errors of 23.3 ms and 14.9 ms in Patients 1 and 2, respectively. Large errors in the SG triggers with respect to ECG would lead to temporal blurring due to mismatch between the true cardiac phase and bin assignment, which may manifest as underestimation of volumetric flow rate and velocity. For both patient scans, the precision error was smaller than the temporal resolution; therefore, the impact of this temporal blurring on flow quantification is not expected to be significant.}

 This work has several limitations. First, while the inversion is made computationally tractable by approximate message passing \cite{Rangan2011}, ReVEAL4D is still computationally intensive and may not be practical in a clinical environment. As demonstrated by Ma et al. \cite{Ma2019}, one solution is to decouple the frequency encode dimension and separately reconstruct each 2D problem in parallel, maximizing available resources at the cost of decreased spatial sparsity. Second, the need to repeatedly acquire self-gating readouts accounts for 11\% (30 seconds) of total acquisition time which could otherwise be spent acquiring additional samples. Decreasing the rate at which self-gating readouts are acquired can improve efficiency but may negatively impact the precision of cardiac trigger detection. \textk{Third, highly variable or irregular heartbeats can decrease the efficiency of our proposed cardiac binning method. For heartbeats longer than the mean, samples in late diastole between the edge of the final ($20^\text{th}$) bin and next trigger point were discarded; for heartbeats shorter than the mean, the final bin or bins were not fully populated, leading to higher acceleration rates for those bins. This approach does not involve compressing or stretching beats to a fixed R-R interval and thus avoids potential flow distortions from beat-to-beat variations in the duration of the diastolic phase. Furthermore, an arrhythmia rejection criterion was applied which discards all samples from heartbeats with R-R intervals substantially deviating from the mean. For instance, a total of seven heartbeats were discarded from Patient 2 due to arrhythmia. While discarding these outlier beats sacrifices efficiency, it offers a substantial robustness to arrhythmia, thereby preserving data consistency and avoiding image corruption in these patients. Fourth, the proposed sampling pattern combined with cardiac binning can lead to repeated sampling of the same indices of the k-space array. In this work, the sample corresponding to the maximum respiratory weight was selected, while the other repeats were discarded. However, all samples could be feasibly included, for example, through weighted averaging of the repeat samples.}  Fifth, the self-gating pipeline applies temporal filtering to initially isolate the cardiac and respiratory surrogate signals, which in our experience, generally improves the robustness of PCA. However, this incorporates specific assumptions regarding heart rate and respiration which may not hold universally across all patients. Adaptive spectral selection as described by Di Sopra et al.\ \cite{DiSopra2019} and singular spectrum analysis \cite{Rosenzweig2018} have both been proposed to automate this process. Sixth, the PCA-driven $v_R$ provides relative relationships between respiratory phases, with absolute displacement obscured through the arbitrary scaling of the right singular vectors. Therefore, the fixed respiratory efficiency of 50\% does not account for individual respiratory patterns, potentially resulting in variability of residual motion artifact between subjects. The cross-correlation method proposed by Han et al. \cite{Han2017} may provide appropriate scaling in millimeters. Finally, at steady-state, the large excitation volume and short TR saturates blood pool magnetization, effectively reducing blood signal magnitude and blood-tissue contrast. While contrast is sufficient for vessel or intra-chamber flow quantification, valve tracking or functional analysis may be challenging. Previous work has explored the use of the exogenous blood pool agent, ferumoxytol, concomitant with 4D flow imaging for contrast enhancement \cite{Cheng2017,KHanneman2016}. Alternatively, 4D flow imaging may be coupled with accelerated, self-gated bSSFP 3D cine to meet application-specific requirements~\cite{DiSopra2019}.

\section{CONCLUSION}
In this work, we have demonstrated a comprehensive whole-heart 4D flow imaging pipeline requiring only five-minutes of total acquisition time to obtain accurate and reliable flow quantification. The straightforward setup and short acquisition time are attractive qualities that may benefit the long-term clinical translation and impact of comprehensive 4D flow examinations.

\section{ACKNOWLEDGMENTS}
This work was funded in part by NIH projects R21EB026657 and R01HL135489. Example MATLAB codes to generate the sampling pattern and implement the processing and reconstruction pipeline, with example datasets, may be found at the following web addresses: \textk{https://github.com/OSU-CMR-AP/Self-Gated-4D-Flow}.

\clearpage
%\bibliography{main_bib.bbl}

\begin{thebibliography}{43}
\providecommand{\natexlab}[1]{#1}
\providecommand{\url}[1]{\texttt{#1}}
\providecommand{\urlprefix}{}

\bibitem[{Nayak et~al.(2015)Nayak, Krishna S and Nielsen, Jon-Fredrik and
  Bernstein, Matt A and Markl, Michael and Gatehouse, Peter D and Botnar, Rene
  M and Saloner, David and Lorenz, Christine and Wen, Han and Hu, Bob S and
  others}]{Nayak2015}
Nayak KS, Nielsen JF, Bernstein MA, Markl M, Gatehouse PD, Botnar RM, et~al.
\newblock Cardiovascular magnetic resonance phase contrast imaging.
\newblock Journal of Cardiovascular Magnetic Resonance 2015;17(1):71.

\bibitem[{da~Silveira et~al.(2017)da Silveira, Juliana Serafim and Smyke,
  Matthew and Rich, Adam V and Liu, Yingmin and Jin, Ning and Scandling, Debbie
  and Dickerson, Jennifer A and Rochitte, Carlos E and Raman, Subha V and
  Potter, Lee C and others}]{DaSilveira2017}
da~Silveira JS, Smyke M, Rich AV, Liu Y, Jin N, Scandling D, et~al.
\newblock Quantification of aortic stenosis diagnostic parameters: comparison
  of fast 3 direction and 1 direction phase contrast CMR and transthoracic
  echocardiography.
\newblock Journal of Cardiovascular Magnetic Resonance 2017;19(1):35.

\bibitem[{Nordmeyer et~al.(2013)Nordmeyer, Sarah and Riesenkampff, Eug{\'e}nie
  and Messroghli, Daniel and Kropf, Siegfried and Nordmeyer, Johannes and
  Berger, Felix and Kuehne, Titus}]{Nordmeyer2013}
Nordmeyer S, Riesenkampff E, Messroghli D, Kropf S, Nordmeyer J, Berger F,
  et~al.
\newblock Four-dimensional velocity-encoded magnetic resonance imaging improves
  blood flow quantification in patients with complex accelerated flow.
\newblock Journal of Magnetic Resonance Imaging 2013;37(1):208--216.

\bibitem[{Markl et~al.(2012)Markl, Michael and Frydrychowicz, Alex and Kozerke,
  Sebastian and Hope, Mike and Wieben, Oliver}]{Markl2012}
Markl M, Frydrychowicz A, Kozerke S, Hope M, Wieben O.
\newblock 4D flow MRI.
\newblock Journal of Magnetic Resonance Imaging 2012;36(5):1015--1036.

\bibitem[{Markl et~al.(2016)Markl, Michael and Schnell, Susanne and Wu, C and
  Bollache, E and Jarvis, K and Barker, Alex J and Robinson, Joshua D and
  Rigsby, Cynthia K}]{Markl2016}
Markl M, Schnell S, Wu C, Bollache E, Jarvis K, Barker AJ, et~al.
\newblock Advanced flow MRI: emerging techniques and applications.
\newblock Clinical Radiology 2016;71(8):779--795.

\bibitem[{Markl et~al.(2010)Markl, Michael and Wallis, Wolf and Brendecke,
  Stefanie and Simon, Jan and Frydrychowicz, Alex and Harloff,
  Andreas}]{Markl2010}
Markl M, Wallis W, Brendecke S, Simon J, Frydrychowicz A, Harloff A.
\newblock Estimation of global aortic pulse wave velocity by flow-sensitive 4D
  MRI.
\newblock Magnetic Resonance in Medicine 2010;63(6):1575--1582.

\bibitem[{Owen and Raptis(2016)Owen, Joseph W and Raptis, Constantine
  A}]{Owen2016}
Owen JW, Raptis CA.
\newblock Emerging Clinical Applications of 4D Flow MR in the Heart and Aorta.
\newblock Current Radiology Reports 2016;4(12):62.

\bibitem[{Vasanawala et~al.(2015)Vasanawala, Shreyas S and Hanneman, Kate and
  Alley, Marcus T and Hsiao, Albert}]{SSVasanawala2015}
Vasanawala SS, Hanneman K, Alley MT, Hsiao A.
\newblock Congenital heart disease assessment with 4D flow MRI.
\newblock Journal of Magnetic Resonance Imaging 2015;42(4):870--886.

\bibitem[{Valverde et~al.(2010)Valverde, Israel and Simpson, John and
  Schaeffter, Tobias and Beerbaum, Philipp}]{Valverde2010}
Valverde I, Simpson J, Schaeffter T, Beerbaum P.
\newblock 4D phase-contrast flow cardiovascular magnetic resonance:
  comprehensive quantification and visualization of flow dynamics in atrial
  septal defect and partial anomalous pulmonary venous return.
\newblock Pediatric Cardiology 2010;31(8):1244--1248.

\bibitem[{Geiger et~al.(2011)Geiger, J and Markl, M and Jung, B and Grohmann, J
  and Stiller, B and Langer, M and Arnold, R}]{Geiger2011}
Geiger J, Markl M, Jung B, Grohmann J, Stiller B, Langer M, et~al.
\newblock 4D-MR flow analysis in patients after repair for tetralogy of Fallot.
\newblock European Radiology 2011;21(8):1651--1657.

\bibitem[{Hope et~al.(2007)Hope, Thomas A and Markl, Michael and Wigstr{\"o}m,
  Lars and Alley, Marcus T and Miller, D Craig and Herfkens, Robert
  J}]{Hope2007}
Hope TA, Markl M, Wigstr{\"o}m L, Alley MT, Miller DC, Herfkens RJ.
\newblock Comparison of flow patterns in ascending aortic aneurysms and
  volunteers using four-dimensional magnetic resonance velocity mapping.
\newblock Journal of Magnetic Resonance Imaging 2007;26(6):1471--1479.

\bibitem[{Westenberg et~al.(2008)Westenberg, Jos JM and Roes, Stijntje D and
  Ajmone Marsan, Nina and Binnendijk, Nico MJ and Doornbos, Joost and Bax,
  Jeroen J and Reiber, Johan HC and De Roos, Albert and van der Geest, Robert
  J}]{JJWestenberg2008}
Westenberg JJ, Roes SD, Ajmone~Marsan N, Binnendijk NM, Doornbos J, Bax JJ,
  et~al.
\newblock Mitral valve and tricuspid valve blood flow: accurate quantification
  with 3D velocity-encoded MR imaging with retrospective valve tracking.
\newblock Radiology 2008;249(3):792--800.

\bibitem[{Pruessmann et~al.(1999)Pruessmann, Klaas P and Weiger, Markus and
  Scheidegger, Markus B and Boesiger, Peter}]{Pruessmann1999}
Pruessmann KP, Weiger M, Scheidegger MB, Boesiger P.
\newblock SENSE: sensitivity encoding for fast MRI.
\newblock Magnetic Resonance in Medicine 1999;42(5):952--962.

\bibitem[{Griswold et~al.(2002)Griswold, Mark A and Jakob, Peter M and
  Heidemann, Robin M and Nittka, Mathias and Jellus, Vladimir and Wang, Jianmin
  and Kiefer, Berthold and Haase, Axel}]{Griswold2002}
Griswold MA, Jakob PM, Heidemann RM, Nittka M, Jellus V, Wang J, et~al.
\newblock Generalized autocalibrating partially parallel acquisitions (GRAPPA).
\newblock Magnetic Resonance in Medicine 2002;47(6):1202--1210.

\bibitem[{Bollache et~al.(2018)Bollache, Emilie and Barker, Alex J and Dolan,
  Ryan Scott and Carr, James C and van Ooij, Pim and Ahmadian, Rouzbeh and
  Powell, Alex and Collins, Jeremy D and Geiger, Julia and Markl,
  Michael}]{Bollache2018}
Bollache E, Barker AJ, Dolan RS, Carr JC, van Ooij P, Ahmadian R, et~al.
\newblock k-t accelerated aortic 4D flow MRI in under two minutes: Feasibility
  and impact of resolution, k-space sampling patterns, and respiratory
  navigator gating on hemodynamic measurements.
\newblock Magnetic Resonance in Medicine 2018;79(1):195--207.

\bibitem[{Ma et~al.(2019)Ma, Liliana E and Markl, Michael and Chow, Kelvin and
  Huh, Hyungkyu and Forman, Christoph and Vali, Alireza and Greiser, Andreas
  and Carr, James and Schnell, Susanne and Barker, Alex J and others}]{Ma2019}
Ma LE, Markl M, Chow K, Huh H, Forman C, Vali A, et~al.
\newblock Aortic 4D flow MRI in 2 minutes using compressed sensing, respiratory
  controlled adaptive k-space reordering, and inline reconstruction.
\newblock Magnetic Resonance in Medicine 2019;81(6):3675--3690.

\bibitem[{Knobloch et~al.(2013)Knobloch, Verena and Boesiger, Peter and
  Kozerke, Sebastian}]{Knobloch2013}
Knobloch V, Boesiger P, Kozerke S.
\newblock Sparsity transform k-t principal component analysis for accelerating
  cine three-dimensional flow measurements.
\newblock Magnetic Resonance in Medicine 2013;70(1):53--63.

\bibitem[{Rich et~al.(2019)Rich, Adam and Potter, Lee C and Jin, Ning and Liu,
  Yingmin and Simonetti, Orlando P and Ahmad, Rizwan}]{Rich2019}
Rich A, Potter LC, Jin N, Liu Y, Simonetti OP, Ahmad R.
\newblock A Bayesian approach for 4D flow imaging of aortic valve in a single
  breath-hold.
\newblock Magnetic Resonance in Medicine 2019;81(2):811--824.

\bibitem[{Ehman and Felmlee(1989)Ehman, Richard L and Felmlee, Joel
  P}]{Ehman1989}
Ehman RL, Felmlee JP.
\newblock Adaptive technique for high-definition MR imaging of moving
  structures.
\newblock Radiology 1989;173(1):255--263.

\bibitem[{Feng et~al.(2016)Feng, Li and Axel, Leon and Chandarana, Hersh and
  Block, Kai Tobias and Sodickson, Daniel K and Otazo, Ricardo}]{Feng2016}
Feng L, Axel L, Chandarana H, Block KT, Sodickson DK, Otazo R.
\newblock XD-GRASP: golden-angle radial MRI with reconstruction of extra
  motion-state dimensions using compressed sensing.
\newblock Magnetic Resonance in Medicine 2016;75(2):775--788.

\bibitem[{Han et~al.(2017)Han, Fei and Zhou, Ziwu and Han, Eric and Gao, Yu and
  Nguyen, Kim-Lien and Finn, J Paul and Hu, Peng}]{Han2017}
Han F, Zhou Z, Han E, Gao Y, Nguyen KL, Finn JP, et~al.
\newblock Self-gated 4D multiphase, steady-state imaging with contrast
  enhancement (MUSIC) using rotating cartesian K-space (ROCK): validation in
  children with congenital heart disease.
\newblock Magnetic Resonance in Medicine 2017;78(2):472--483.

\bibitem[{Feng et~al.(2018)Feng, Li and Coppo, Simone and Piccini, Davide and
  Yerly, Jerome and Lim, Ruth P and Masci, Pier Giorgio and Stuber, Matthias
  and Sodickson, Daniel K and Otazo, Ricardo}]{Feng2018}
Feng L, Coppo S, Piccini D, Yerly J, Lim RP, Masci PG, et~al.
\newblock 5D whole-heart sparse MRI.
\newblock Magnetic Resonance in Medicine 2018;79(2):826--838.

\bibitem[{Winkelmann et~al.(2006)Winkelmann, Stefanie and Schaeffter, Tobias
  and Koehler, Thomas and Eggers, Holger and Doessel, Olaf}]{Winkelmann2006}
Winkelmann S, Schaeffter T, Koehler T, Eggers H, Doessel O.
\newblock An optimal radial profile order based on the Golden Ratio for
  time-resolved MRI.
\newblock IEEE Transactions on Medical Imaging 2006;26(1):68--76.

\bibitem[{Lustig et~al.(2007)Lustig, Michael and Donoho, David and Pauly, John
  M}]{Lustig2007}
Lustig M, Donoho D, Pauly JM.
\newblock Sparse MRI: The application of compressed sensing for rapid MR
  imaging.
\newblock Magnetic Resonance in Medicine 2007;58(6):1182--1195.

\bibitem[{Larson et~al.(2004)Larson, Andrew C and White, Richard D and Laub,
  Gerhard and McVeigh, Elliot R and Li, Debiao and Simonetti, Orlando
  P}]{Larson2004}
Larson AC, White RD, Laub G, McVeigh ER, Li D, Simonetti OP.
\newblock Self-gated cardiac cine MRI.
\newblock Magnetic Resonance in Medicine 2004;51(1):93--102.

\bibitem[{Stehning et~al.(2005)Stehning, C and B{\"o}rnert, P and Nehrke, K and
  Eggers, H and Stuber, M}]{Stehning2005}
Stehning C, B{\"o}rnert P, Nehrke K, Eggers H, Stuber M.
\newblock Free-breathing whole-heart coronary MRA with 3D radial SSFP and
  self-navigated image reconstruction.
\newblock Magnetic Resonance in Medicine 2005;54(2):476--480.

\bibitem[{Uribe et~al.(2009)Uribe, Sergio and Beerbaum, Philipp and
  S{\o}rensen, Thomas Sangild and Rasmusson, Allan and Razavi, Reza and
  Schaeffter, Tobias}]{Uribe2009}
Uribe S, Beerbaum P, S{\o}rensen TS, Rasmusson A, Razavi R, Schaeffter T.
\newblock Four-dimensional (4D) flow of the whole heart and great vessels using
  real-time respiratory self-gating.
\newblock Magnetic Resonance in Medicine 2009;62(4):984--992.

\bibitem[{Cheng et~al.(2017)Cheng, Joseph Y and Zhang, Tao and Alley, Marcus T
  and Uecker, Martin and Lustig, Michael and Pauly, John M and Vasanawala,
  Shreyas S}]{Cheng2017}
Cheng JY, Zhang T, Alley MT, Uecker M, Lustig M, Pauly JM, et~al.
\newblock Comprehensive multi-dimensional MRI for the simultaneous assessment
  of cardiopulmonary anatomy and physiology.
\newblock Scientific Reports 2017;7(1):1--15.

\bibitem[{Lustig et~al.(2007)Lustig, M and Cunningham, CH and Daniyalzade, E
  and Pauly, JM}]{Lustig2007b}
Lustig M, Cunningham C, Daniyalzade E, Pauly J.
\newblock Butterfly: a self navigating Cartesian trajectory.
\newblock In: Proceedings of the 15th Annual Meeting of ISMRM, Berlin, Germany;
  2007. p. 865.

\bibitem[{Bastkowski et~al.(2018)Bastkowski, Rene and Weiss, Kilian and Maintz,
  David and Giese, Daniel}]{Bastkowski2018}
Bastkowski R, Weiss K, Maintz D, Giese D.
\newblock Self-gated golden-angle spiral 4D flow MRI.
\newblock Magnetic Resonance in Medicine 2018;80(3):904--913.

\bibitem[{Walheim et~al.(2019)Walheim, Jonas and Dillinger, Hannes and Kozerke,
  Sebastian}]{Walheim2019}
Walheim J, Dillinger H, Kozerke S.
\newblock Multipoint 5D flow cardiovascular magnetic resonance-accelerated
  cardiac-and respiratory-motion resolved mapping of mean and turbulent
  velocities.
\newblock Journal of Cardiovascular Magnetic Resonance 2019;21(1):42.

\bibitem[{Cheng et~al.(2013)Cheng, JY and Uecker, M and Alley, MT and
  Vasanawala, SS and Pauly, JM and Lustig, M}]{Cheng2013}
Cheng J, Uecker M, Alley M, Vasanawala S, Pauly J, Lustig M.
\newblock Free-breathing pediatric imaging with nonrigid motion correction and
  parallel imaging.
\newblock In: Proceedings of the 21st Annual Meeting of ISMRM; 2013. p. 312.

\bibitem[{Rich et~al.(2016)Rich, Adam and Potter, Lee C and Jin, Ning and Ash,
  Joshua and Simonetti, Orlando P and Ahmad, Rizwan}]{Rich2016}
Rich A, Potter LC, Jin N, Ash J, Simonetti OP, Ahmad R.
\newblock AB ayesian model for highly accelerated phase-contrast MRI.
\newblock Magnetic Resonance in Medicine 2016;76(2):689--701.

\bibitem[{Rangan(2011)Rangan, Sundeep}]{Rangan2011}
Rangan S.
\newblock Generalized approximate message passing for estimation with random
  linear mixing.
\newblock In: 2011 IEEE International Symposium on Information Theory
  Proceedings IEEE; 2011. p. 2168--2172.

\bibitem[{Liu et~al.(2008)Liu, Bo and Zou, Yi Ming and Ying, Leslie}]{Liu2008}
Liu B, Zou YM, Ying L.
\newblock SparseSENSE: application of compressed sensing in parallel MRI.
\newblock In: 2008 International Conference on Information Technology and
  Applications in Biomedicine IEEE; 2008. p. 127--130.

\bibitem[{Ting et~al.(2017)Ting, Samuel T and Ahmad, Rizwan and Jin, Ning and
  Craft, Jason and Serafim da Silveira, Juliana and Xue, Hui and Simonetti,
  Orlando P}]{ting2017fast}
Ting ST, Ahmad R, Jin N, Craft J, Serafim~da Silveira J, Xue H, et~al.
\newblock Fast implementation for compressive recovery of highly accelerated
  cardiac cine MRI using the balanced sparse model.
\newblock Magnetic Resonance in Medicine 2017;77(4):1505--1515.

\bibitem[{Walsh et~al.(2000)Walsh, David O and Gmitro, Arthur F and Marcellin,
  Michael W}]{Walsh2000}
Walsh DO, Gmitro AF, Marcellin MW.
\newblock Adaptive reconstruction of phased array MR imagery.
\newblock Magnetic Resonance in Medicine 2000;43(5):682--690.

\bibitem[{Heiberg et~al.(2010)Heiberg, Einar and Sj{\"o}gren, Jane and Ugander,
  Martin and Carlsson, Marcus and Engblom, Henrik and Arheden,
  H{\aa}kan}]{Heiberg2010}
Heiberg E, Sj{\"o}gren J, Ugander M, Carlsson M, Engblom H, Arheden H.
\newblock Design and validation of Segment-freely available software for
  cardiovascular image analysis.
\newblock BMC Medical Imaging 2010;10(1):1.

\bibitem[{Pruitt et~al.(2019)Pruitt, Aaron A and Jin, Ning and Liu, Yingmin and
  Simonetti, Orlando P and Ahmad, Rizwan}]{Pruitt2019}
Pruitt AA, Jin N, Liu Y, Simonetti OP, Ahmad R.
\newblock A method to correct background phase offset for phase-contrast MRI in
  the presence of steady flow and spatial wrap-around artifact.
\newblock Magnetic Resonance in Medicine 2019;81(4):2424--2438.

\bibitem[{Cheng et~al.(2015)Cheng, Joseph Y and Zhang, Tao and
  Ruangwattanapaisarn, Nichanan and Alley, Marcus T and Uecker, Martin and
  Pauly, John M and Lustig, Michael and Vasanawala, Shreyas S}]{cheng2015free}
Cheng JY, Zhang T, Ruangwattanapaisarn N, Alley MT, Uecker M, Pauly JM, et~al.
\newblock Free-breathing pediatric MRI with nonrigid motion correction and
  acceleration.
\newblock Journal of Magnetic Resonance Imaging 2015;42(2):407--420.

\bibitem[{Di~Sopra et~al.(2019)Di Sopra, Lorenzo and Piccini, Davide and Coppo,
  Simone and Stuber, Matthias and Yerly, J{\'e}r{\^o}me}]{DiSopra2019}
Di~Sopra L, Piccini D, Coppo S, Stuber M, Yerly J.
\newblock An automated approach to fully self-gated free-running cardiac and
  respiratory motion-resolved 5D whole-heart MRI.
\newblock Magnetic Resonance in Medicine 2019;82(6):2118--2132.

\bibitem[{Rosenzweig et~al.(2018)Rosenzweig, Sebastian and Scholand, Nick and
  Holme, H Christian M and Uecker, Martin}]{Rosenzweig2018}
Rosenzweig S, Scholand N, Holme HCM, Uecker M.
\newblock Cardiac and Respiratory Self-Gating in Radial MRI using an Adapted
  Singular Spectrum Analysis (SSA-FARY).
\newblock arXiv preprint arXiv:181209057 2018;.

\bibitem[{Hanneman et~al.(2016)Hanneman, Kate and Kino, Aya and Cheng, Joseph Y
  and Alley, Marcus T and Vasanawala, Shreyas S}]{KHanneman2016}
Hanneman K, Kino A, Cheng JY, Alley MT, Vasanawala SS.
\newblock Assessment of the precision and reproducibility of ventricular
  volume, function, and mass measurements with ferumoxytol-enhanced 4D flow
  MRI.
\newblock Journal of Magnetic Resonance Imaging 2016;44(2):383--392.

\end{thebibliography}

%
%
\clearpage
\section{FIGURES AND TABLES}

\newpage

\newpage

\newpage

\newpage

\clearpage
\newpage

\newpage

\newpage

\newpage

% Table 1

\newpage
%\section{TABLES}
\begin{table}[htb]
\centering
	\begin{tabular}{@{}lccc@{}}
		\toprule
		\textbf{Subjects}                        & \multicolumn{3}{c}{Eight healthy subjects, age range $20 - 46$, 2 female} \\ 
		
		\vspace{2mm}\textbf{Scanner}                         & \multicolumn{3}{c}{MAGNETOM Prisma (3T)}                                                                                                 \\		
				& {\hspace{5mm}\bf{SG 4D Flow}}                                         & {\hspace{15mm}\bf{2D-PC}}        & {\hspace{10mm}\bf{4D-GRAPPA}} \\
				\cline{2-4}

		\textbf{FOV {[}mm{]}}                    & \multicolumn{1}{c}{}                                                           &              &                                          \\
		\hspace{3 mm}\textit{\textbf{Frequency}}              & \hspace{5mm}$220 - 250$       & \hspace{15mm}$300 - 400$      & \hspace{10mm}$360 - 420$                                  \\
		\hspace{3 mm}\textit{\textbf{Phase}}                  & \hspace{5mm}$220 - 250$       & \hspace{15mm}$160 - 320$      & \hspace{10mm}$215 - 260$                                  \\
		\hspace{3 mm}\textit{\textbf{Slice}}                  & \hspace{5mm}$129 - 146$       & \hspace{15mm}N/A              & \hspace{10mm}$46 - 52$                                    \\
		\textbf{Spatial Res. {[}mm{]}}     &                                                                                &              &                                          \\
		\hspace{3 mm}\textit{\textbf{Frequency}}              & \hspace{5mm}\multirow{3}{*}{\begin{tabular}[c]{@{}c@{}}\vspace{-2mm}$2.3 - 2.6$\\ (Isotropic)\end{tabular}} & \hspace{15mm}$1.9 - 2.5$      & \hspace{10mm}$2.3 - 2.6$                                  \\
		\hspace{3 mm}\textit{\textbf{Phase}}                  &                   & \hspace{15mm}$2.5 - 3.3$      & \hspace{10mm}$3.0 - 3.5$                                    \\
		\hspace{3 mm}\textit{\textbf{Slice}}                  &                                                                                & \hspace{15mm}6.0            & \hspace{10mm}$2.3 - 2.6$                                  \\
		\textbf{Temporal Res. {[}ms{]}}    & \hspace{5mm}$43 - 65$ \hspace{0.25 mm}$^{\dagger}$         & \hspace{15mm}$43 - 45$        & \hspace{10mm}38                                       \\
		\textbf{TE {[}ms{]}}                     & \hspace{5mm}2.2                                            & \hspace{15mm}$2.4 - 2.6$      & \hspace{10mm}2.3                                      \\
		\textbf{TR {[}ms{]}}                     & \hspace{5mm}4.0                                            & \hspace{15mm}$4.3 - 4.5$      & \hspace{10mm}4.8                                      \\
		\textbf{Flip Angle {[}degrees{]}}        & \hspace{5mm}7                                                                              & \hspace{15mm}15           & \hspace{10mm}7                                        \\
		\textbf{Receiver BW {[}Hz/px{]}}  & \hspace{5mm}801                                                                            & \hspace{15mm}560          & \hspace{10mm}495                                      \\
		\textbf{VENC {[}cm/s{]}}                 & \hspace{5mm}150                                                                            & \hspace{15mm}150          & \hspace{10mm}150                                      \\
		\textbf{Respiratory Eff. {[}\%{]}} & \hspace{5mm}50\hspace{0.25 mm}$^{\dagger}$                                                                             & \hspace{15mm}Breath-held  & \hspace{10mm}$31 - 79$                                    \\
		\textbf{Acquisition Time}                & \hspace{5mm}5 min                                         & \hspace{15mm}$7 - 14$ s       & \hspace{10mm}$4.6 - 14.1$ min                         \\
		\textbf{Acceleration Rate}               & \hspace{5mm}$19 - 20$ \hspace{0.25 mm}$^{\dagger}$                                                                         & \hspace{15mm}GRAPPA,  2   & \hspace{10mm}GRAPPA, 3                                \\ \bottomrule
	\end{tabular}
	\caption{MR acquisition parameters \textk{in healthy subjects for the proposed self-gated 4D flow (SG 4D Flow), conventional 2D phase contrast (2D-PC), and conventional navigator-gated 4D flow (4D-GRAPPA) protocols}. Ranges indicate minimum and maximum values. \textk{Parameters given for} 2D-PC represent the aggregate of all vessels/planes imaged. $^{\dagger}$Indicated values were determined retrospectively after data acquisition. }
	\label{tab:scan_param}
\end{table}

\newpage
\begin{table}[h]
\centering
\begin{tabular}{lcccc}
\\ \toprule
                                                    & \multicolumn{2}{c}{\textbf{Patient 1}}            & \multicolumn{2}{c}{\textbf{Patient 2}}        \\ \hline

\vspace{-2mm}\textbf{Description}                                   & \multicolumn{2}{l}{37-year-old male with a}        &
\multicolumn{2}{l}{80-year-old male with coronary}\\

                                                    \vspace{-2mm}& \multicolumn{2}{l}{history of congenital}        &
\multicolumn{2}{l}{artery disease and premature}\\

                                                    \vspace{-2mm}& \multicolumn{2}{l}{pulmonary artery stenosis.}        &
\multicolumn{2}{l}{ventricular contraction burden.}\\ \vspace{-4mm}\\

\textbf{Scanner}                                    & \multicolumn{4}{c}{MAGNETOM Sola (1.5T)} \\ \vspace{-12mm} \\

\textbf{}                                           & \multicolumn{2}{c}{}                              & \multicolumn{2}{c}{}                          \\
\textbf{}                                           & \textbf{SG 4D Flow}       & \textbf{2D-PC}\hspace{0.25 mm}$^{*}$        & \textbf{SG 4D Flow}   & \textbf{2D-PC}    \\ \cline{2-5} 
\textbf{FOV {[}mm{]}}                               &                           &                       &                       &                   \\
\hspace{3 mm}\textit{\textbf{Frequency}}            & 300                       & 380                   & 320                   & 380               \\
\hspace{3 mm}\textit{\textbf{Phase}}                & 300                       & 285                   & 320                   & 285               \\
\hspace{3 mm}\textit{\textbf{Slice}}                & 179                       & N/A                   & 185                   & N/A               \\
\textbf{Spatial Res. {[}mm{]}}                      &                           &                       &                       &                   \\
\hspace{3 mm}\textit{\textbf{Frequency}}            & 3.1                       & 2.0                   & 3.3                   & 2.0               \\
\hspace{3 mm}\textit{\textbf{Phase}}                & 3.1                       & 2.9                   & 3.3                   & 2.9               \\
\hspace{3 mm}\textit{\textbf{Slice}}                & 3.2                       & 6.0                   & 3.3                   & 6.0               \\
\textbf{Temporal Res. {[}ms{]}}                     & 45\hspace{0.25 mm}$^{\dagger}$                        & 42                    & 65\hspace{0.25 mm}$^{\dagger}$                    & 118$^{\triangle}$               \\
\textbf{TE {[}ms{]}}                                & 2.5                       & 2.3                   & 2.6                   & 2.3               \\
\textbf{TR {[}ms{]}}                                & 4.4                       & 4.2                   & 4.6                   & 4.2               \\
\textbf{Flip Angle {[}degrees{]}}                   & 7                         & 15                    & 7                     & 15                \\
\textbf{Receiver BW {[}Hz/px{]}}                    & 801                       & 501                   & 801                   & 501               \\
\textbf{VENC {[}cm/s{]}}                            & 200                       & $150 - 200$              & 300                   & 150               \\
\textbf{Respiratory Eff. {[}\%{]}}                  & 50\hspace{0.25 mm}$^{\dagger}$                        & Breath-held           & 50\hspace{0.25 mm}$^{\dagger}$                    & Breath-held       \\
\textbf{Acquisition Time}                           & 5 min                     & $8 - 9$ s                & 5 min                 & 12 s              \\
\textbf{Acceleration Rate}                          & 23\hspace{0.25 mm}$^{\dagger}$                      & GRAPPA, 2               & 26\hspace{0.25 mm}$^{\dagger}$                  & GRAPPA, 2    
\\ \bottomrule
\end{tabular}
\caption{\textk{MR acquisition parameters used in the patient study for the proposed self-gated 4D flow (SG 4D Flow) and conventional 2D phase contrast (2D-PC) protocols. $^{*}$Ranges indicate minimum and maximum values from the aggregate of all vessels imaged. $^{\dagger}$Values were determined retrospectively after data acquisition. $^{\triangle}$A lower temporal resolution was used to avoid corruption by cardiac arrhythmia in this patient. The images were, however, reconstructed on a temporal grid with 20 cardiac bins.}}
\label{tab:patient_parameters}
\end{table}

% Table 2
\newpage
\begin{table}[h!]
\footnotesize
\begin{tabular}{lccccccccc}
\toprule
\rowcolor[HTML]{EFEFEF} 
\multicolumn{1}{c}{\cellcolor[HTML]{EFEFEF}}          & \multicolumn{9}{c}{\cellcolor[HTML]{EFEFEF}\textbf{Patient 1}}                                                                                                                                                                                                                                                                                                                                                                  \\ \hline

                                                      \vspace{-1mm}& \multicolumn{3}{c}{\bf{2D-PC}}                                                                                                                                                                      & \multicolumn{3}{c}{\bf{L1-SENSE}}                                                                                                                                                                   & \multicolumn{3}{c}{\bf{ReVEAL4D}}                                                                                                                                                                   \\
                                                      & \begin{tabular}[c]{@{}c@{}}\vspace{-2mm}Net Flow \\ {[}mL{]}\end{tabular} & \begin{tabular}[c]{@{}c@{}}\vspace{-2mm}Pk. Vel. \\ {[}cm/s{]}\end{tabular} & \begin{tabular}[c]{@{}c@{}}\vspace{-2mm}Pk. Flow \\ {[}mL/s{]}\end{tabular} & \begin{tabular}[c]{@{}c@{}}\vspace{-2mm}Net Flow \\ {[}mL{]}\end{tabular} & \begin{tabular}[c]{@{}c@{}}\vspace{-2mm}Pk. Vel. \\ {[}cm/s{]}\end{tabular} & \begin{tabular}[c]{@{}c@{}}\vspace{-2mm}Pk. Flow \\ {[}mL/s{]}\end{tabular} & \begin{tabular}[c]{@{}c@{}}\vspace{-2mm}Net Flow \\ {[}mL{]}\end{tabular} & \begin{tabular}[c]{@{}c@{}}\vspace{-2mm}Pk. Vel. \\ {[}cm/s{]}\end{tabular} & \begin{tabular}[c]{@{}c@{}}\vspace{-2mm}Pk. Flow \\ {[}mL/s{]}\end{tabular} \\ \cline{2-10} 
\textbf{Aao}                                          & 91.4                                                         & 106.6                                                          & 442.2                                                          & 91.3                                                         & 88.1                                                           & 459.5                                                          & 86.8                                                         & 95.3                                                           & 450.8                                                          \\
\textbf{MPA}                                          & 65.8                                                         & 161.1                                                          & 428.5                                                          & 60.1                                                         & 152.9                                                          & 420.5                                                          & 62.6                                                         & 161.3                                                          & 427.7                                                          \\
\textbf{RPA}                                          & 8.3                                                          & 74.9                                                           & 29.9                                                           & 3.5                                                          & 56.4                                                           & 34.2                                                           & 8.8                                                          & 73.3                                                           & 51.9                                                           \\
\textbf{LPA}                                          & 68.0                                                         & 99.4                                                           & 308.6                                                          & 64.4                                                         & 79.6                                                           & 297.5                                                          & 73.4                                                         & 92.8                                                           & 322.4                                                          \\ \hline
\rowcolor[HTML]{EFEFEF} 
\multicolumn{1}{c}{\cellcolor[HTML]{EFEFEF}\textbf{}} & \multicolumn{9}{c}{\cellcolor[HTML]{EFEFEF}\textbf{Patient 2}}                                                                                                                                                                                                                                                                                                                                                                                                                                                                                                                                   \\ \hline
\multicolumn{1}{c}{\textbf{Aao}}                      & 92.6                                                         & 109.4                                                          & 395.4                                                          & 93.8                                                         & 78.8                                                           & 391.4                                                          & 92.0                                                         & 110.0                                                          & 405.6   \\
\bottomrule
\end{tabular}
%\bottomrule
\caption{\textk{Summary of flow quantification in \textk{patients.} \textk{Mean errors from L1-SENSE/ReVEAL4D with respect to 2D-PC for each flow index are: -2.6/-0.5 mL (net flow), -19.1/-3.8 cm/s (peak velocity), and -0.3/10.8 mL/s (peak flow rate).} Note that only the Aao was scanned using 2D-PC in the second patient. Here, peak through-plane velocity and peak volumetric flow rate are abbreviated as Pk. Vel. and Pk. Flow, respectively.}}
\label{tab:patient_flow}
\end{table}

\newpage
\clearpage

\end{document}